\documentclass[onecolumn,preprintnumbers,amsmath,amssymb,nofootinbib,11pt]{revtex4}
\pdfoutput=1
\usepackage{epsfig}
\usepackage{CJK}
\usepackage{amsfonts}
\usepackage{graphicx}
\usepackage{epsfig}
\usepackage{eepic}
\usepackage{amsmath}
\usepackage{amssymb}
\usepackage{xcolor}
\usepackage{bbm}
\usepackage{dcolumn}
\usepackage{bm}
\usepackage{ulem}
\usepackage{slashbox}

\usepackage{hyperref}
\hypersetup{colorlinks=false}

\def\bc{\begin{center}}
\def\nno{\nonumber}
\def\ec{\end{center}}
\def\be{\begin{eqnarray}}
\def\ee{\end{eqnarray}}
\definecolor{dyellow}{rgb}{1.,0.8,.0}
\definecolor{myblue}{rgb}{.1,.1,.7}
\definecolor{dcyan}{rgb}{.0,.6,.6}
\definecolor{dmagenta}{rgb}{0.6,0.0,0.6}
\definecolor{brown}{rgb}{0.6,0.2,0.}
\definecolor{darkblue}{rgb}{.0,.0,0.5}
\definecolor{darkred}{rgb}{0.75,0.0,0.0}
\definecolor{orange}{rgb}{1.,.6,.0}
\definecolor{dorange}{rgb}{0.8,.4,.0}
\definecolor{darkgreen}{rgb}{0.0,0.6,0.0}
\definecolor{purple}{rgb}{.4,.0,.4}
\definecolor{lightgrey}{rgb}{0.7, 0.7, 0.7}
\definecolor{grey}{rgb}{0.4, 0.4, 0.4}

\def\ga{\gamma}

\def\pa{\partial}

\newcommand{\nc}{\newcommand}
\nc{\rnc}{\renewcommand} \nc{\ket}[1]{\left | \, #1 \right \rangle}
\nc{\bra}[1]{\left \langle #1 \, \right |}
\nc{\ua}{\uparrow} \nc{\da}{\downarrow}

\nc{\braket}[2]{\langle\, #1\,|\,#2\,\rangle}
\nc{\half}{\frac{1}{2}}

\nc{\prj}{\mathcal{P}} \nc{\hilb}{\mathcal{H}}
\nc{\pth}{\mathcal{C}} \nc{\inprod}[2]{\braket{#1}{#2}}
\nc{\upket}{\ket{\uparrow}} \nc{\downket}{\ket{\downarrow}}
\nc{\upbra}{\bra{\uparrow}} \nc{\downbra}{\bra{\downarrow}}

\begin{document}
\begin{CJK*}{GBK}{song}

\hspace{11.4cm}\eprint{CCTP-2014-30}

\hspace{11.4cm}\eprint{CCQCN-2014-58}

\title{Time Evolution of Entanglement Entropy in Quenched Holographic Superconductors}
\author{Xiaojian Bai$^1$} \email{xiaojian.bai@physics.gatech.edu}
\author{Bum-Hoon Lee$^2$} \email{bhl@sogang.ac.kr}
\author{Li Li$^3$} \email{lili@physics.uoc.gr}
\author{Jia-Rui Sun$^4$} \email{jrsun@ecust.edu.cn}
\author{Hai-Qing Zhang$^5$}\email{h.q.zhang@uu.nl}

\affiliation{${}^1$   School of Physics, Georgia Institute of Technology, Atlanta, GA 30332, USA;}
\affiliation{${}^2$ Center for Quantum Spacetime, Sogang University, Seoul 121-742, Republic of Korea, \\
\& Department of Physics, Sogang University, Seoul 121-742, Republic of Korea;}
\affiliation{${}^3$ Crete Center for Theoretical Physics, Department of Physics, University of Crete, 71003 Heraklion, Greece;}
\affiliation{${}^4$ Department of Physics and Institute of Modern Physics, East China University of Science and Technology, Shanghai 200237, China;}
\affiliation{${}^5$ Institute for Theoretical Physics, Utrecht University, Leuvenlaan 4, 3584 CE Utrecht, The Netherlands}

\begin{abstract}
We investigate the dynamical evolution of entanglement entropy in a holographic superconductor model by quenching the source term of the dual charged scalar operator. By access to the full background geometry, the holographic entanglement entropy is calculated for a strip geometry at the AdS boundary. It is found that the entanglement entropy exhibits a robust non-monotonic behaviour in time, independent of the strength of Gaussian quench and the size of the strip: it first displays a small dip, then grows linearly, and finally saturates. In particular, the linear growth velocity of the entanglement entropy has an upper bound for strip with large width; The equilibrium value of the non-local probe at late time shows a power law scaling behaviour with respect to the quench strength; Moreover, the entanglement entropy can uncover the dynamical transition at certain critical quench strength which happens to coincide with the one obtained form the dynamical evolution of scalar order parameter.
\end{abstract}

\maketitle

 \newpage
 \tableofcontents

\section{Introduction}

As a measurement of quantum entanglement in given systems,  entanglement entropy is a fundamental notion in quantum physics. For example, in condensed matter physics it can be used to probe the quantum phase transitions near the critical point \cite{Calabrese:2004eu,Calabrese:2009qy}. Considering a subregion $\bold{{A}}$ with the remaining parts as $\bold{{B}}$, the von Neumann entanglement entropy of $\bold{{A}}$ is
$
S_{\bold{A}}=-\rm{tr}_{\bold{A}}(\rho_{\bold{A}}\ln(\rho_{\bold{A}})),
$
in which $\rho_{\bold{A}}=\rm{tr}_{\bold B}\rho$ is the reduced density matrix for the subregion $\bold A$ by tracing out the degrees of freedom of $\bold B$, while $\rho$ is the density matrix of the whole system. Since entanglement entropy can measure the entanglement between subsystems,  for a dynamical  situation, the entanglement entropy can describe the evolution of quantum entanglement in the system.  Some interesting results have been obtained in~\cite{Calabrese:2005in} for $(1+1)$-dimension, in which some powerful results are available by using the techniques in conformal field theory (CFT). It was conjectured that entanglement propagating from small to large scales has an allowed maximum  speed which is taken to be one. However, a natural extension to study the evolution of entanglement entropy for field theories in higher dimensions is very complicated. So far, few results are available from field theory approach.

In the AdS/CFT correspondence \cite{Maldacena:1997re}, the holographic entanglement entropy (HEE) was conjectured by Ryu and Takayanagi \cite{Ryu:2006bv} that the entanglement entropy of the subregion $\bold A$ can be given by the following area law,
\be\label{HEE}
S_{\rm HEE}=\frac{\rm Area(\gamma_{\bold A})}{4 G_N},
\ee
in which $\gamma_{\rm A}$ is the minimal surface which ends at the border of subregion $\bold A$ on the AdS boundary, and $G_N$ is the Newton's constant of the bulk gravity. The Ryu-Takayanagi proposal for the holographic derivation of entanglement entropy in static background was soon generalized to time dependent situations~\cite{Hubeny:2007xt} and  has been generically proven by~\cite{Lewkowycz:2013nqa}. This proposal provides an elegant and executable way to calculate entanglement entropy of a strongly coupled system which has a gravity dual.
Applications of the formula \eqref{HEE} have been extended to a bunch of models, for reviews one can refer to \cite{Takayanagi:2012kg}.

In particular, taking advantage of this holographic formula, time evolution of entanglement entropy has been studied recently in general space-time dimensions~\cite{AbajoArrastia:2010yt,Albash:2010mv,Balasubramanian:2010ce,Balasubramanian:2011ur,Allais:2011ys,Keranen:2011xs,Caceres:2012em,Nozaki:2013wia,Hartman:2013qma,Nozaki:2013vta,Liu:2013iza,Li:2013sia,Liu:2013qca,Hubeny:2013dea,Abajo-Arrastia:2014fma, Asplund:2014coa,Buchel:2014gta,Caceres:2014pda}. It was uncovered that there is a common linear growth of HEE before it saturates into the final equilibrium states. Actually, this linear growth of HEE is reminiscent of the linear behaviour of entanglement entropy by field theory calculation~\cite{Calabrese:2005in} in $(1+1)$-dimension. More precisely, the authors of~\cite{Calabrese:2005in} showed that the entanglement entropy for a segment of width $2L$ grows with time linearly as $\Delta S(t, L)=s_{\rm eq}2L~t$ before saturation. In the above relation, $\Delta S$ represents the difference between the entanglement entropy from that at the initial time, while $s_{\rm eq}$ is the equilibrium thermal entropy density. This linear relation was also analytically studied in holographic aspects by the authors of~\cite{Liu:2013iza,Liu:2013qca}, in which the analogous velocity $v_E$ was introduced by the formula
\be
\Delta S_{\Sigma}(t)=s_{\rm eq} A_{\Sigma}v_E t,
\ee
with $A_{\Sigma}$ the ``area" of subregion $\Sigma$ one considered. They also speculated that the linear growth velocity had an upper bound  for general dimensions.~\footnote{In the case of a segment with width $2L$, $A_{\Sigma}=2L$, which is consistent with the CFT calculation~\cite{Calabrese:2005in} .}

The behaviour of entanglement entropy for static case in some holographic superconductor (superfluid)
models has been studied in~\cite{Albash:2012pd,Cai:2012sk,Cai:2012nm,Cai:2012es,Li:2013rhw,Kuang:2014kha,Yao:2014fua,Peng:2014ira}.
It turns out that the entanglement entropy is a good probe to investigate the holographic phase transitions. It can
indicate not only the appearance, but also the order of the phase transition.
In this paper, we would like to generalise previous study to  out of equilibrium situation, i.e., to consider the dynamical evolution of HEE in holographic superconductors.
In particular, we adopt the basic model for the time evolution of holographic superconductor from~\cite{Bhaseen:2012gg}.~\footnote{ For other studies on dynamical evolution of holographic superconductors, readers can consult~\cite{Murata,Gao:2012aw,Garcia-Garcia:2013rha,Bai:2014poa,Sonner:2014tca,Chesler:2014gya,Alishahiha:2014cwa}.} Depending on the temperature of the system, the equilibrium state can be described by an AdS Reissner-Nordstorm (AdS-RN) black hole or a hairy black hole, corresponding to a ``normal" phase or a superconducting phase of the dual field theory, respectively.
The authors of~\cite{Bhaseen:2012gg} quenched the system by a boundary scalar source of charged order parameter in terms of a Gaussian type quench, and then they found three distinct regimes for the order parameters depending on the strength of the quench. In our paper, after reproducing the numerical results which perfectly match~\cite{Bhaseen:2012gg}, we study the dynamical HEE with a strip geometry on this background. We would like to study how this non-local observable evolves compared to the scalar order parameter and whether HEE can be used as a probe to detect different patterns of {dynamical symmetry breaking}.

The Gaussian quench is localised at a particular time, say $t=0$.  The quench has two parameters, one controls its amplitude and the other one controls the speed of quench.  In order to compare with the results in~\cite{Bhaseen:2012gg}, we fix the later parameter as well and to study how the quench strength  affects the HEE.  Based on the explicit numerical calculation, we uncover some universal features of  HEE as a function of time, which is independent of the strength of quench.
When the quench time is very near $t=0$, HEE as a function of time develops a small dip. The depth and slope of the dip depend on the quench strength. Specifically, as quench is stronger the depth and slope of the dip will be larger. The physics behind this dip is still vague to us.
As time goes beyond this dip, HEE will perform a linear growth until it saturates to a particular value which depends on the quench strength and size of the strip subsystem. We find that HEE will grow more rapidly if the quench is stronger during the linear regime. At the saturation time, we find that there exists a continuous saturation if the width of the strip is relatively short; however, if the width is long enough, we find a swallow tail of HEE at the saturation time. This swallow tail behaviour can be understood by looking at multiple minimal area surfaces near the saturation time. However, for the entanglement entropy we need to choose the minimal area surface.
After the saturation time, HEE will enter the equilibrium states. We find that for a fixed width of the strip, stronger quenches will induce larger values of HEE at equilibrium. Physically, this can be understood that stronger quench will pump more energy or degrees of freedom into the system.

We also analyse the time evolution of the event horizon and apparent horizon. We see that at late time they coincide together, since now it is in the equilibrium state. For large width of the strip, the tip point of the minimal surface will also meet the horizons above, which can be easily understood from a geometry point of view: as the strip width is large enough the minimal surface probes deeply into the bulk. Nevertheless, before the equilibrium time, the location of the event horizon, apparent horizon and the tip point of the minimal surface will not meet together. Especially, evolving from the initial time, the tip point will cross the event horizon first, and then at late time it meets the horizon. We extract the linear growth velocity $v_E$ and the entanglement entropy density at final equilibrium states as a function of quench strength. Both quantities increase with the quench strength and there is a particular quench strength where the behaviour of two quantities changes qualitatively. Interestingly, this critical strength is the one that exceeding which the charged order parameter evolves into the normal states with an over-damped behaviour, which was studied in \cite{Bhaseen:2012gg} already. The saturated entanglement entropy density at late time versus quench strength exhibits power law behaviour but with different powers as the quench is smaller or larger than the critical value.

This paper is arranged as follows: In Section \ref{sect:basic}, we introduce the basic holographic setup; We will reproduce the numerical results of the dynamical holographic superconductor after quench in Section \ref{sect:hsc}; Time evolution of HEE will be shown in Section \ref{sect:HEE}; In Section \ref{sect:pthee}, we use the HEE to probe the phase transition which is studied in section \ref{sect:hsc}; Finally, we draw our conclusions and discussions in Section \ref{sect:con}.

\section{Holographic Setup}
\label{sect:basic}
 We adopt the 4-dimensional Einstein-Maxwell-complex scalar action as the model,
 \be\label{action}
 \mathfrak{S}=\int d^4x\sqrt{-g}\left[R+\frac{6}{\ell^2}-\frac14F_{\mu\nu}F^{\mu\nu}-|\pa_\mu\psi-iqA_\mu\psi|^2-m^2|\psi|^2\right],
 \ee
 in which $A_\mu$ is the Maxwell gauge field dual to the conserved current in the boundary
identified as the weakly-gauged electromagnetic field in the context of superconductors, $F_{\mu\nu}=\pa_\mu A_\nu-\pa_\nu A_\mu$ is the field strength,  $\psi$ is the complex scalar field and $\ell$ is the radius of the AdS spacetime.  We will consider the time dependent and spatially homogenous and isotropic black hole background, its general form is like the Vaidya form,
 \be\label{metric}
 ds^2=\frac{1}{z^2}\left[-F(t,z)dt^2-2dtdz+S(t,z)^2(dx^2+dy^2)\right],
 \ee
in which $t$ is the ingoing Eddington-Finkelstein time coordinate while $z$ is the radial direction in the bulk and $z=0$ is the infinite boundary.  For the static case, the Hawking temperature of the black hole is well defined, which reads
\be\label{temperature}
T=-\frac{1}{4\pi}\frac{dF}{dz}\bigg{|}_{z=z_h},
\ee
where the event horizon $z_h$ is the minimum zero point of $F(z)=0$.
In order to construct a model of holographic superconductor (superfluid), we make the ansatz of the fields as usual \cite{Bhaseen:2012gg},
\be\label{ansatz}
\psi=\psi(t,z), ~~ A=(A_t(t,z),0,0,0).
\ee

Therefore, from the action and the metric one can derive the equations of motion (EoMs) for the system \footnote{Readers can refer to  \cite{Murata} for details of the EoMs. One should note that in \cite{Murata} the metric function $\Phi(t,z)$ equals to $S(t,z)/z$ in our paper.},
\be\label{eoms}
 0&=&\left(\frac{S}{z}D(S/z)\right)'-\frac{S^2}{4z^4}\left(\frac12z^4A_t'^2+m^2\psi\psi^*-6\right),\\
 0&=&2z^2(D\psi)'+iqz^2A_t'\psi+\frac{2z^3}{S}\left(D(S/z)\right)\psi'+\frac{2z^3}{S}(S/z)'D\psi+m^2\psi,\\
 0&=&\left(z^2(F/z^2)'\right)'-z^2A_t'^2+\frac{4z^2}{S^2}(D(S/z))(S/z)'-({\psi^*}'D\psi+\psi'D\psi^*),\\
 0&=&2z^2(DA_t)'+z^4(F/z^2)'A_t'+\frac{4z^3}{S}(D(S/z))A_t'-2iq(\psi D\psi^*-\psi^*D\psi),
 \ee
and there are three constraint equations,
\be\label{constraint}
-2D^2(S/z)-(F'-2F/z)D(S/z)-\frac{S}{z}D\psi D\psi^*&=&0,\\
-2\left(z(S/z)''+2(S/z)'\right)-S\psi'{\psi^*}'&=&0,\\
z^2A_t''+2zA_t'+\frac{2z^3}{S}(S/z)'A_t'+iq(\psi{\psi^*}'-\psi^*\psi')&=&0.
\ee
In above expressions a prime $'$ denotes the derivative with respect to $z$ and $D(S/z)=\partial_t(S/z)-\frac F2\partial_z(S/z), DA_t=\partial_t A_t-\frac F2\partial_zA_t, D\psi=\partial_t\psi-\frac F2\partial_z\psi-iqA_t\psi, D^2(S/z)=\partial_t(D(S/z))-\frac F2\partial_z(D(S/z))$.

The expansions of the fields near the boundary $z=0$ can be obtained as (we have set $m^2=-2/\ell^2$ in the following context),
\be\label{expansion}
\psi(t,z)&\sim& z\psi_1(t)+z^2\psi_2(t)+\cdots,~~A_t(t,z)\sim \mu(t)-z\rho(t)+\cdots,\nno\\
F(t,z)&\sim&1-\frac12z^2\left(|\psi_1(t)|^2\right)+\cdots,~~S(t,z)\sim1-\frac14z^2|\psi_1(t)|^2+\cdots.
\ee
From the AdS/CFT dictionary, we can regard $\psi_1$ as the source of dual scalar operator in the boundary CFT; $\psi_2$ is related to the expectation value of the dual operator, explicitly $\langle O(t)\rangle=\psi_2(t)+(iq\mu-\partial_t )\psi_1(t)$; \footnote{In \cite{Bhaseen:2012gg}, the authors have missed a term $\partial_t \psi_1(t)$ for the expectation value $\langle O(t)\rangle$, thanks to Toby Wiseman's private communication.} $\mu$ is the chemical potential and $\rho$ corresponds to the charge density $J^t$ in the boundary CFT by $J^t(t)=\rho(t)-\partial_t\mu(t)$. We shall fix the inner boundary at $z=1$ which is always behind the apparent horizon, so we do not need to impose any specific boundary condition there.

\section{Holographic superconductor after quench}
\label{sect:hsc}

In this section we will briefly review the results in~\cite{Bhaseen:2012gg} we also reproduce perfectly here. The details of the numerical calculations can be found in~\cite{Bhaseen:2012gg} and~\cite{Murata}. Ref.~\cite{Bhaseen:2012gg} adopted the Gaussian type quench centered at $t=0$ as,
\be\label{quench}
\psi_1(t)=(\mu_i\delta) e^{-\left(\frac{t}{\tau/\mu_i}\right)^2}.
\ee
where $\delta$ and $\tau$ are the dimensionless quench strength and width respectively, while $\mu_i$ is the chemical potential at the initial time. From the holographic renormalization one can find that the explicit form of the expectation value for the scalar operator is
\be\label{vev}
\langle O(t)\rangle=\psi_2+(iq\mu-\partial_t )\psi_1=\psi_2+iq\left(-\frac{\text{Im}(\psi_2)}{q\psi_1}\right)\psi_1-\partial_t\psi_1=\text{Re}(\psi_2)-\partial_t\psi_1.
\ee
in which we have chosen the gauge $\mu=-\text{Im}(\psi_2)/(q\psi_1)$ in order to satisfy  $\partial_t\rho=0$ for the whole time. ~\footnote{One should notice that in~\cite{Murata} the authors chose  a gauge by setting the chemical potential $\mu=0$.}

In the numerics, certain parameters are fixed as $q=2$, AdS radius $\ell=1$ and $\tau=0.5$. At the initial hairy state, the temperature is $T_i=0.5T_c$ with $T_c\approx0.090\mu_c$ is the critical temperature when the normal AdS-RN black hole turns unstable to become a hairy black hole with $\psi\neq0$. We make use of the Chebyshev spectral method \cite{Trefethen} in the radial direction while using the 3-step Adams-Bashforth method in the time direction. In practice we start the initial data from $t=-5$ and then let the system evolve in time, while the quench \eqref{quench} is centered at $t=0$, until it reaches the equilibrium state at late time.
\begin{figure}[h]
\includegraphics[trim=3cm 0cm 3cm 0cm, clip=true, scale=.27]{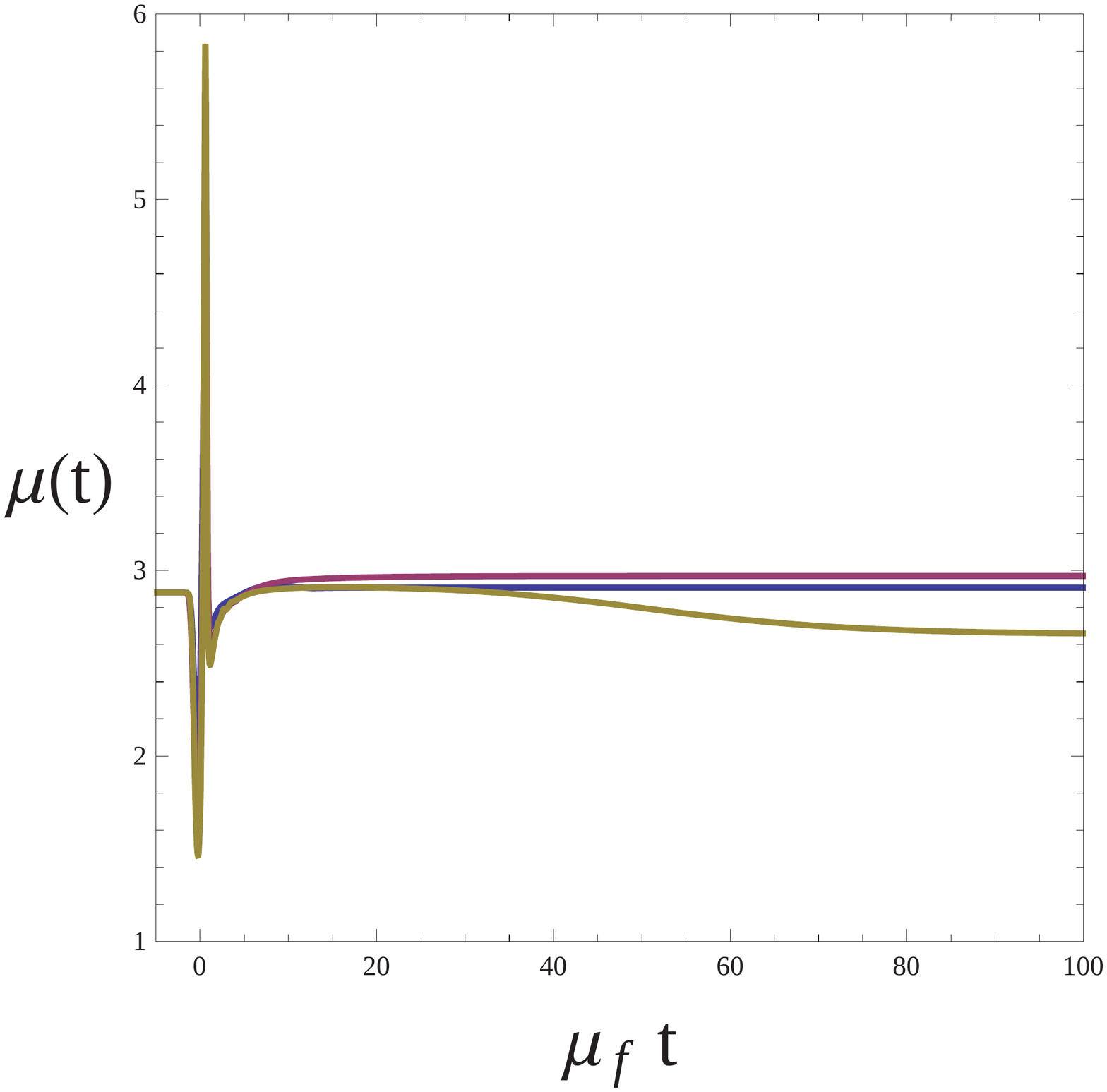}
\includegraphics[trim=0cm 3.cm 0cm 2.8cm, clip=true, scale=.33]{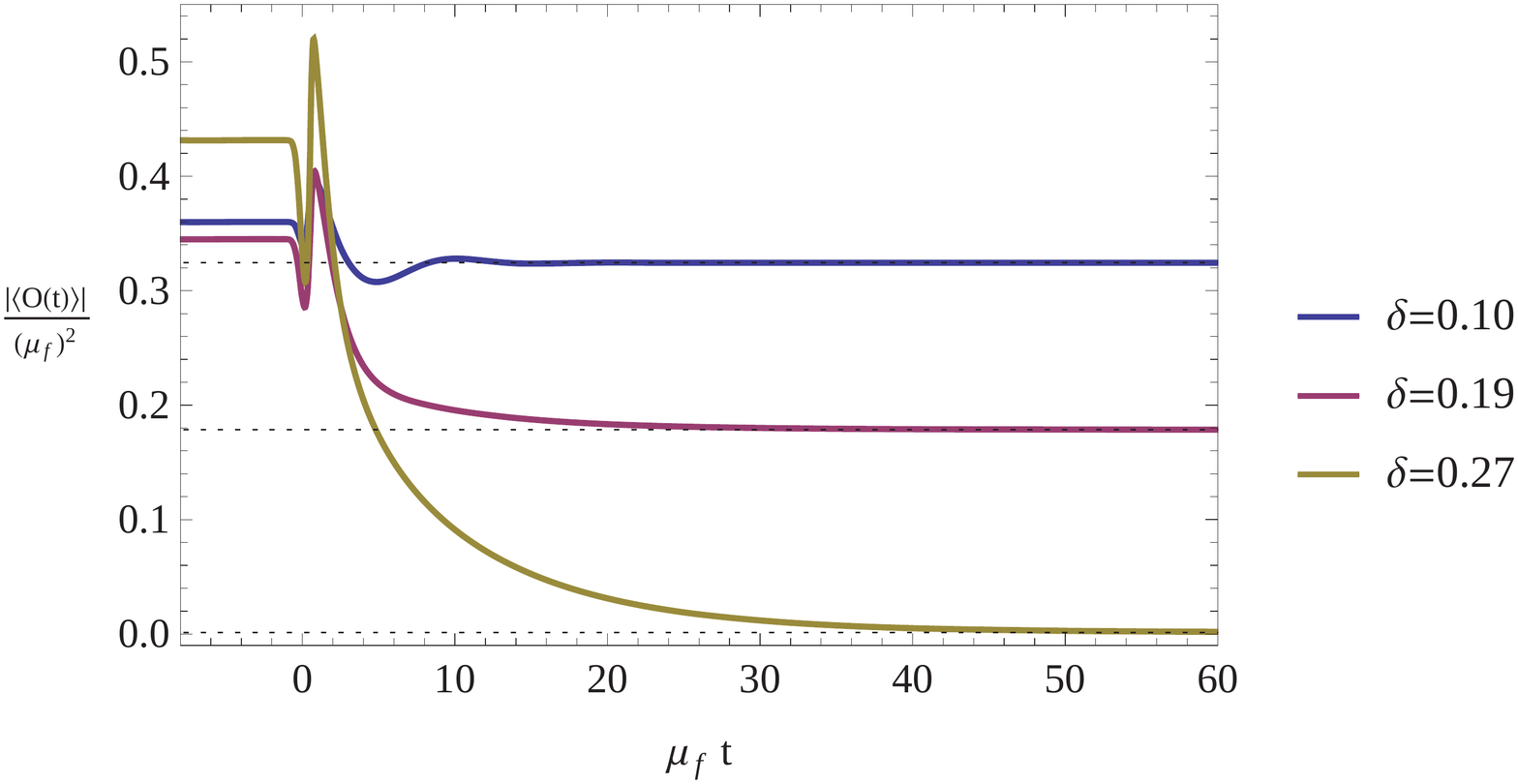}
\caption{\label{Ot}(Left) Time evolution of chemical potential with various quench strengths $\delta$, in which $\mu_f$ is the chemical potential in equilibrium state at late time; (Right) Time evolution of the dimensionless order parameter in three difference regimes with various $\delta$. Dashed lines correspond to the equilibrium values of the order parameter at late time.}
\end{figure}

By varying the quench strength $\delta$, the authors in~\cite{Bhaseen:2012gg} discovered three distinct regimes of the order parameter which can be found in the right panel of Fig.\ref{Ot}. Specifically, when $\delta=0.1$ the order parameter decays oscillatorily to a finite value; However, if one strengthens the quench a little bit, say $\delta=0.19$, the order parameter will only exhibit a damping decay to a finite value without oscillations; Further strengthening the quench to $\delta=0.27$ one only observes a damping order parameter to a vanishing value. On the left panel of Fig.\ref{Ot}, we plot the evolution of the chemical potential for various $\delta$, in which $\mu_f$ is the value of chemical potential in equilibrium state at late time. It shows that $\mu_f(\delta=0.19)>\mu_f(\delta=0.10)>\mu_f(\delta=0.27)$, which exhibits a subtle relation between the quench strength and the equilibrium chemical potential, although the initial states for them are identical.

\begin{figure}[h]
\includegraphics[trim=0cm 0cm 0cm 0cm, clip=true, scale=0.27]{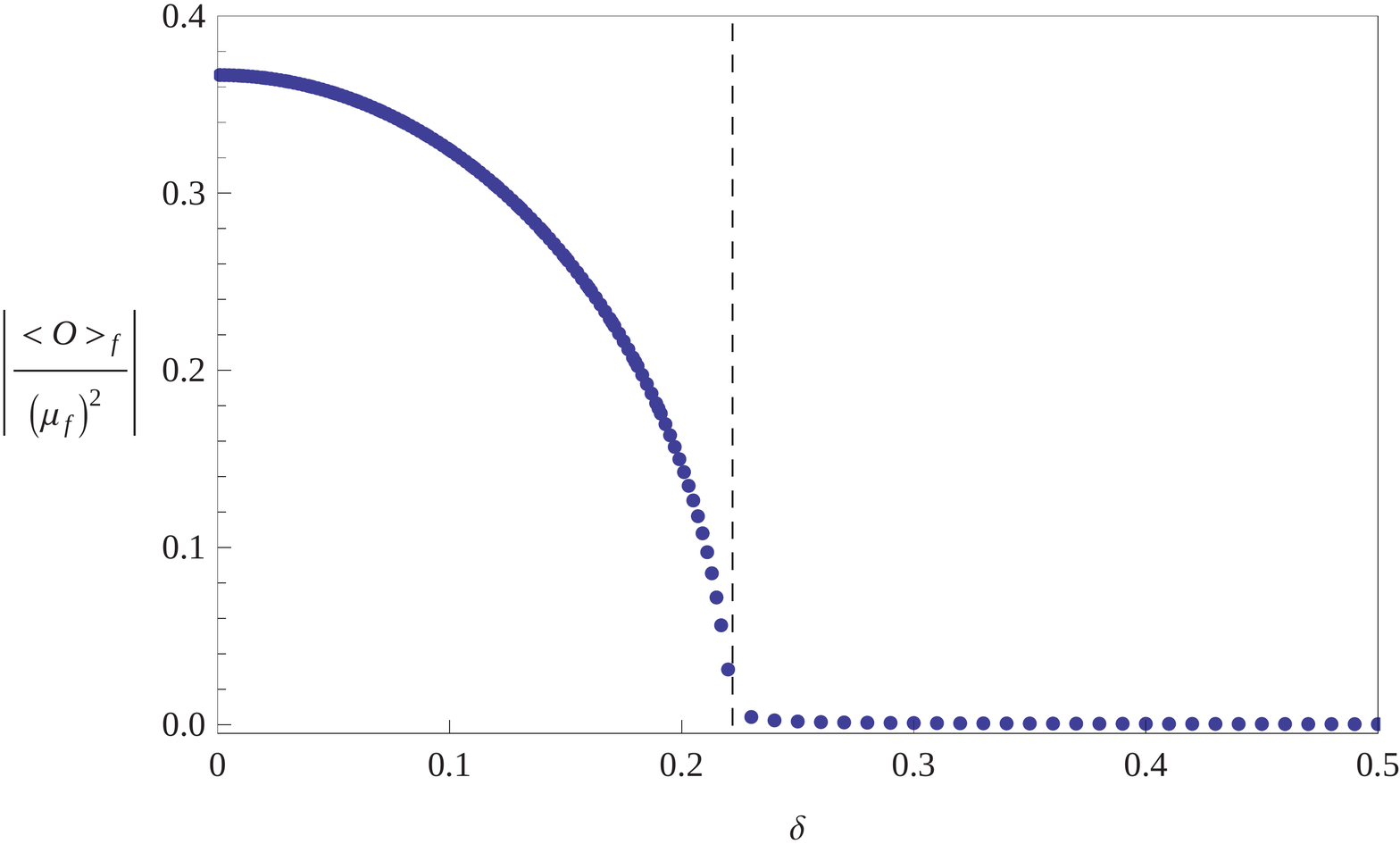}
\includegraphics[trim=0cm 0cm 0cm 0cm, clip=true, scale=0.26]{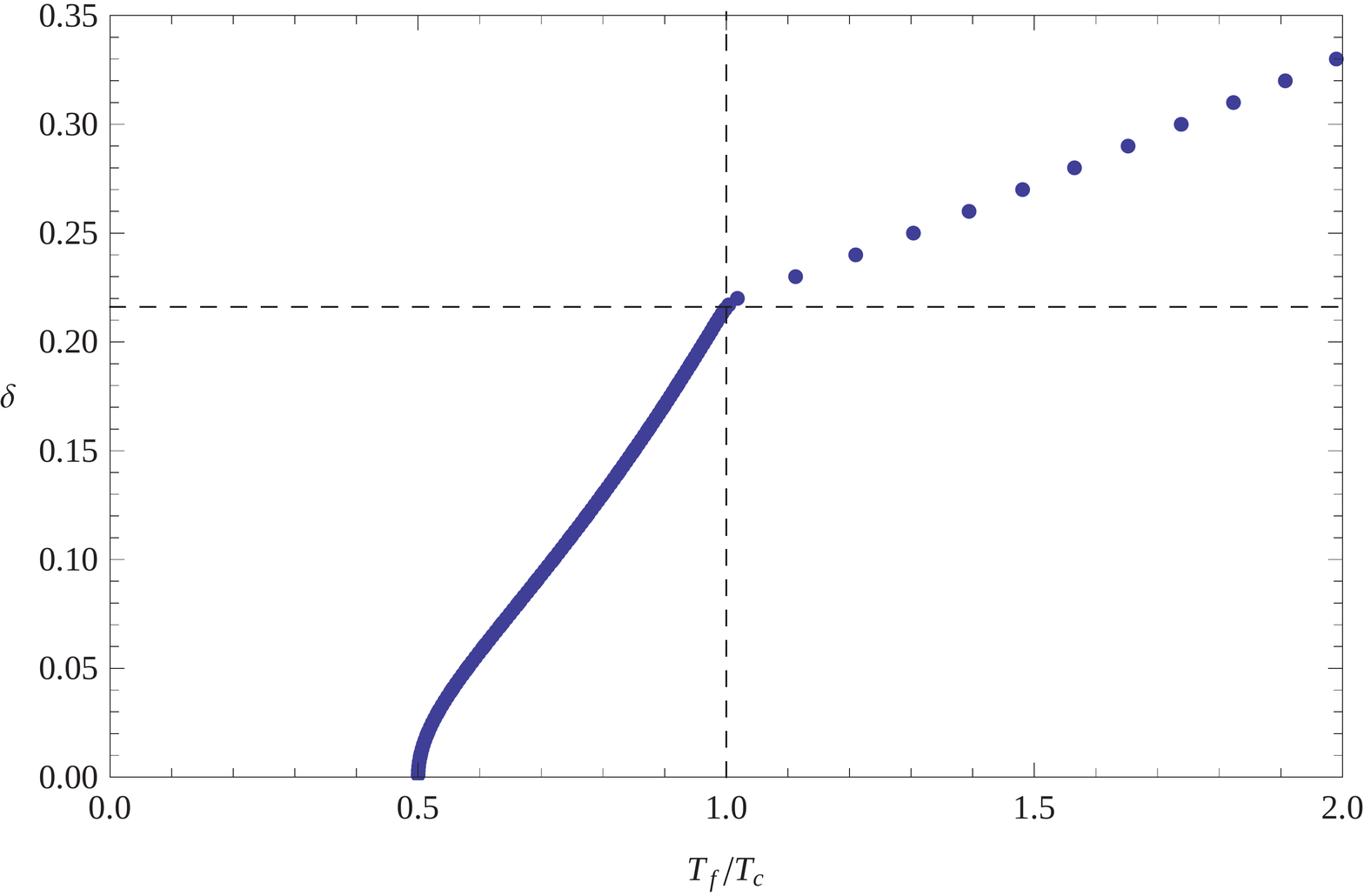}
\caption{\label{fig2a} (Left) The expectation value of operator at equilibrium state versus the quench strength; (Right) The relations between the quench strength and the ratios of the equilibrium temperature over the critical temperature. The dashed lines correspond to the critical point from a finite $\langle O\rangle_f$ to a vanishing $\langle O\rangle_f$. }
\end{figure}
On the left panel of Fig.\ref{fig2a}, we plot the expectation value of the dual charged operator at equilibrium $\langle O\rangle_f$ versus the quench strength $\delta$. It shows that at around $\delta\approx0.22$ there is a phase transition from finite $\langle O\rangle_f$ to vanishing $\langle O\rangle_f$. Actually this critical point corresponds to the one that the final equilibrium temperature $T_f$ identical to the critical temperature $T_c$ in the static case, which is shown on the right panel of Fig.\ref{fig2a}. \footnote{\label{footnote}We do not intend to mention about the other `dynamical' transition point $\delta_*$ or $T_*$ in \cite{Bhaseen:2012gg}, which is defined from the oscillatory damping to un-oscillatory damping order parameter, since the context we showed in this section is enough for our analysis of the HEE in the next section.}

\section{Entanglement entropy in holographic superfluid after quench}
\label{sect:HEE}

\begin{figure}[h]
\includegraphics[trim=5cm 3cm 5.5cm 7cm, clip=true, scale=0.4]{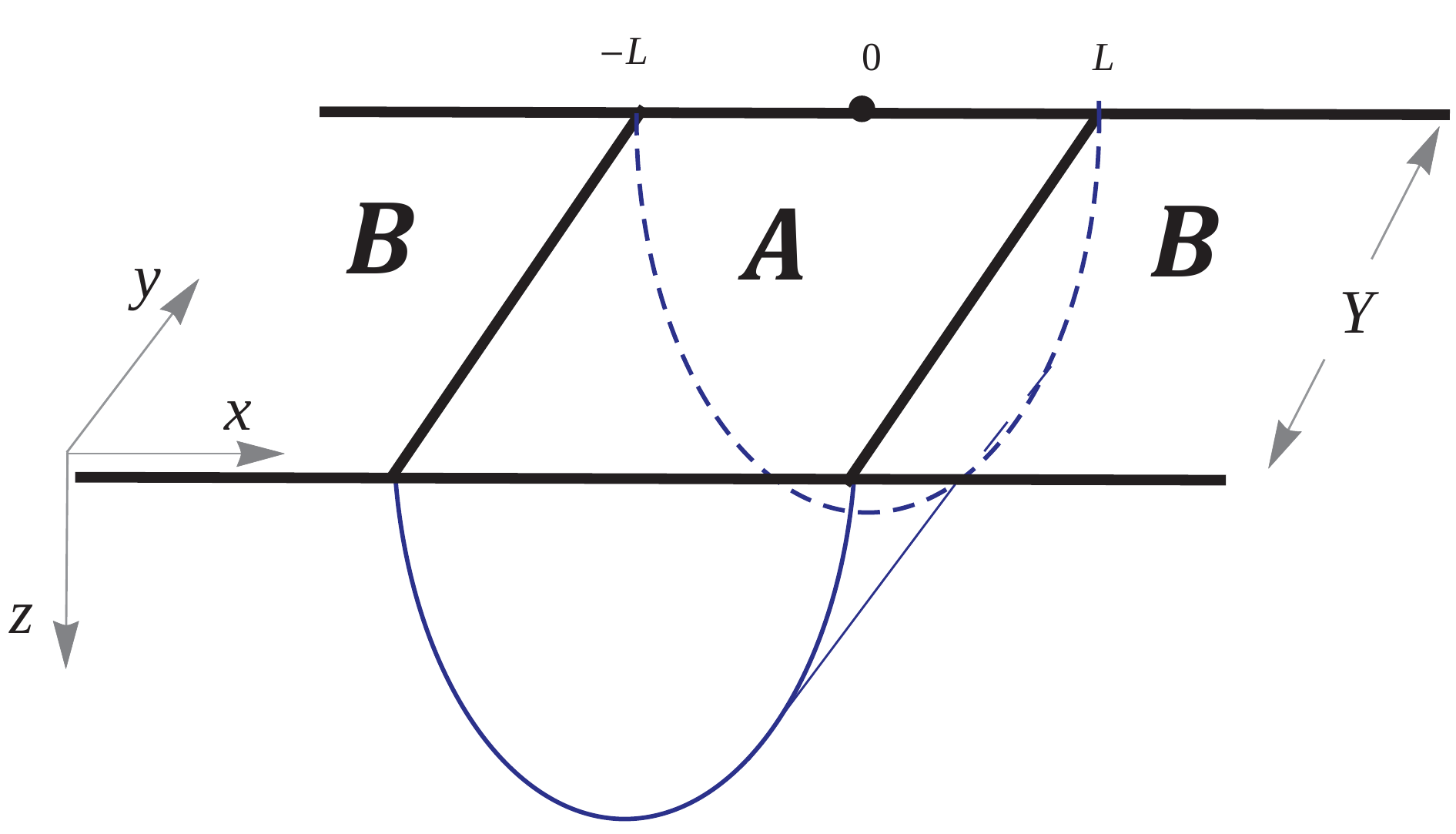}
\caption{\label{strip} Strip $\bold{A}$ with width $2L$ along $x$-direction and length $Y$ along $y$-direction on the boundary; The sketchy blue part is the minimal surface $\gamma_{\bold{A}}$ corresponding to $\bold{A}$. }
\end{figure}
After the preparation for the time evolution of the holographic superconductor in the preceding section, we can now compute the HEE based on this background. The entanglement entropy depends on the choice of the subsystem. In this paper we will focus on a strip geometry which has width $2L$ in $x$ direction and extends in $y$ direction. Please consult Fig.\ref{strip} for details.

By considering  the symmetry of the minimal surface, we can regard $t$ and $z$ as functions of $x$. The holographic dual surface $\gamma_\bold{A}$ is defined by the following embedding,
\be
t=t(x),\quad z=z(x),\quad -\frac{Y}{2}<y<\frac{Y}{2}~ (Y\rightarrow\infty),
\ee
where $Y$ is the regularized length in $y$ direction. We further require that the centre of the strip is located at $x=0$.

This dual surface extends all the way into the bulk. We are interested in the case that the surface is smooth, thus at the tip of the surface one obtains
 \be\label{bdy}
 t(x=0)=t_*, \quad z(x=0)=z_*,  ~~{\rm{and}}~~~ t'|_{x=0}=z'|_{x=0}=0,
 \ee
in which $'$ is the derivative with respect to $x$ and the last relation can be understood since $x=0$ is the middle part of the strip. One should note that the time $t$ in metric \eqref{metric} is the ingoing time in the Eddington-Finkelstein coordinates, for an observer on the boundary the physical time for himself/herself is $t_{\rm{phys}}=t+z$ near $z\to0$. 
Moreover, since HEE will diverge when calculated at the AdS boundary $z=0$, which corresponds to the UV divergence in dual field theory. To regularise this divergence we choose the UV boundary at a cut-off $\epsilon\sim0$ in practice. Thus, the boundary conditions for $t$ and $z$ at $x=L$ are
\be\label{bdy2}
t(x=L)=t_{\rm{phys}}-\epsilon,~~ z(x=L)=\epsilon.
 \ee
The area of the minimal surface $\gamma_{\bold{A}}$ can be obtained as,
\be\label{areaA}
\text{Area}(\ga_{\bold{A}})=Y\int^L_{-L}dx \frac{S\big(t(x),z(x)\big)}{z(x)^2}\sqrt{S\big(t(x),z(x)\big)^2-F\big(t(x),z(x)\big)t'(x)^2-2t'(x)z'(x)}.
\ee
where $t$ and $z$ are functions of $x$, therefore $S$ and $F$ are also functions of $x$.

We can regard the integrand in \eqref{areaA} as a Lagrangian $\mathcal{L}\big(t(x),z(x);t'(x),z'(x)\big)$ with $x$ direction thought of as ``time". Since the Lagrangian does not explicitly depend on ``time" $x$, the Hamiltonian $\mathcal{H}$ is a conserved quantity as $x$ changes,
\be\label{hamilton}
\mathcal{H}=\mathcal{L}-t'\frac{\partial\mathcal{L}}{\partial t'}-z'\frac{\partial\mathcal{L}}{\partial z'}
=\frac{S^3}{z^2\sqrt{S^2-Ft'^2-2t'z'}}\equiv\rm{const.}
\ee
From the boundary conditions at the tip point $z_*$ \eqref{bdy}, we can reach
\be
\sqrt{S^2-Ft'^2-2t'z'}=\frac{z_*^2S^3}{z^2S_*^2},
\ee
where $S_*=S(t_*, z_*)$.
Therefore, we can compute the HEE of strip $\bold{A}$ as,
\be
\label{heestrip}\text{Area}(\ga_{\bold{A}})=2Y\int_0^{L}dx \left(\frac{S^4}{z^4}\frac{z_*^2}{S_*^2}\right)\Rightarrow S_{\rm{HEE}}=\frac{\text{Area}(\ga_{\bold{A}})}{4G_N}.
\ee
In addition, from the Euler-Lagrange equations we can derive that $t(x)$ and $z(x)$ satisfy the EoMs
\be
\label{tz1}\left(S^2-Ft'^2-2t'z'\right)\frac{\pa_tS}{S^3}+\frac{2S\pa_tS-\pa_tFt'^2}{2S^2}+\frac{d}{dx}\left(\frac{Ft'+z'}{S^2}\right)&=&0,\\
\label{tz2}\left(\frac{\pa_zS}{S^3}-\frac{2}{S^2z}\right)\left(S^2-Ft'^2-2t'z'\right)+\frac{2S\pa_zS-\pa_zFt'^2}{2S^2}+\frac{d}{dx}\left(\frac{t'}{S^2}\right)&=&0.
\ee
 Functions $S(t,z)$ and $F(t,z)$ have been already obtained from the preceding section, therefore, we only need  to solve $t(x)$ and $z(x)$ from the above EoMs together with the boundary conditions \eqref{bdy} and \eqref{bdy2}. We take advantage of the shooting method to compute the area of the minimal surface $\gamma_{\bold{A}}$. Moreover, we should also subtract the diverging term from HEE which is proportional to $1/\epsilon$. The scheme we used for the subtraction of the regularized HEE is
\be\label{reghee}
\Delta S_{\rm{HEE}}=S_{\rm{HEE}}-S_{\rm{HEE}}(t_i).
\ee
where $t_i$ is the initial time $t=-5$ as mentioned in the previous section.
In the numerics, we have set $4G_N=\ell=1$ and $\epsilon=10^{-4}$. We will present the numerical results for $z_*$ and $\Delta S_{\rm{HEE}}$ for various $\delta$ and $L$ in the following context.

In our numerical scheme, we fix the strip size $L$ and try to find the minimal surface at each time $t_{\rm{phy}}$. In practice, we compute the minimal surface by first choosing $t_{*(0)}$ which is in the $t$-coordinate at the tip point. Then we adopt the shooting method to find a suitable $z_*$ which satisfies the EoMs \eqref{tz1} and \eqref{tz2} and the boundary conditions \eqref{bdy} and \eqref{bdy2}. More precisely, one finds a particular $z_*$, such that $z(x=L)=\epsilon$. Meanwhile, we can read the value of $t_{\rm{phys}}$ at $x=L$ form those solutions. It is in this way that we get the information of the minimal surface at chosen $t_{*(0)}$. Next, we continue computing the minimal surface for another tip time $t_{*(1)}$ in the same way, and so on so forth. Eventually, we can obtain minimal surfaces, i.e., HEE for all time.

\subsection{Early time dynamics}

\begin{figure}[h!]
$L=0.5$\\
\includegraphics[trim=0cm 0cm 0cm 0cm, clip=true, scale=0.26]{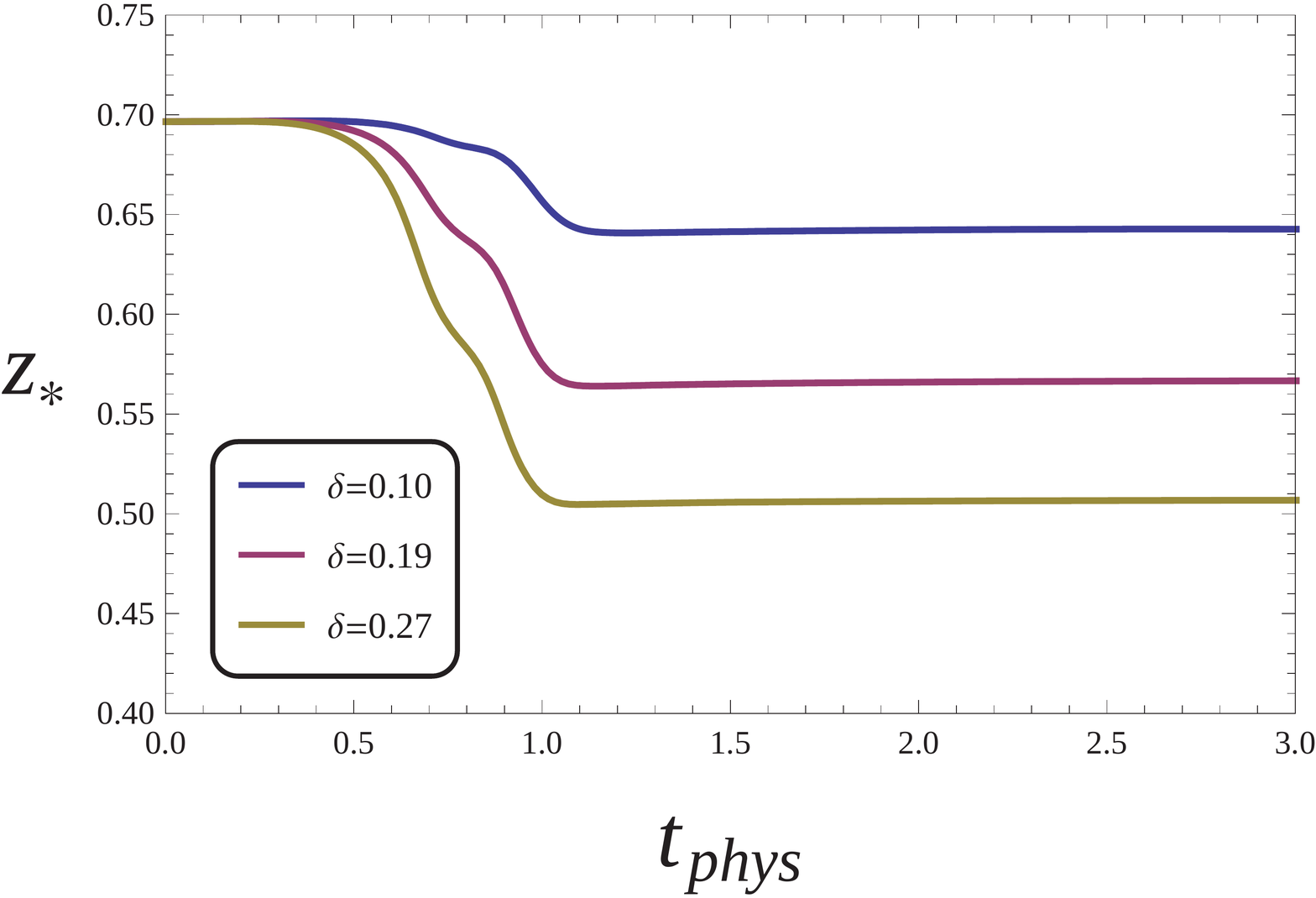}
\includegraphics[trim=0cm 0cm 0cm 0cm, clip=true, scale=0.26]{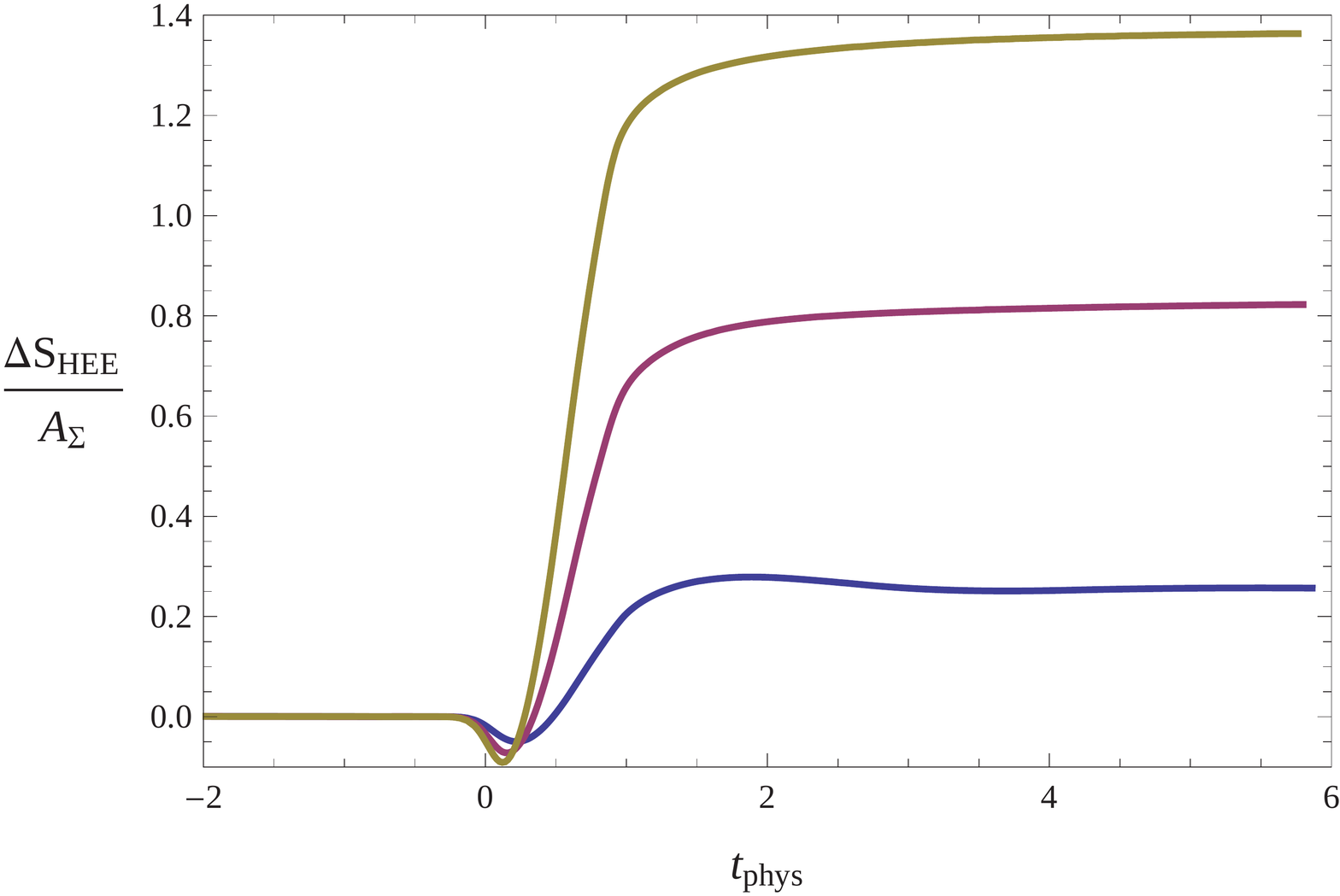}\\
$L=1.0$\\
\includegraphics[trim=0cm 0cm 0cm 0cm, clip=true, scale=0.26]{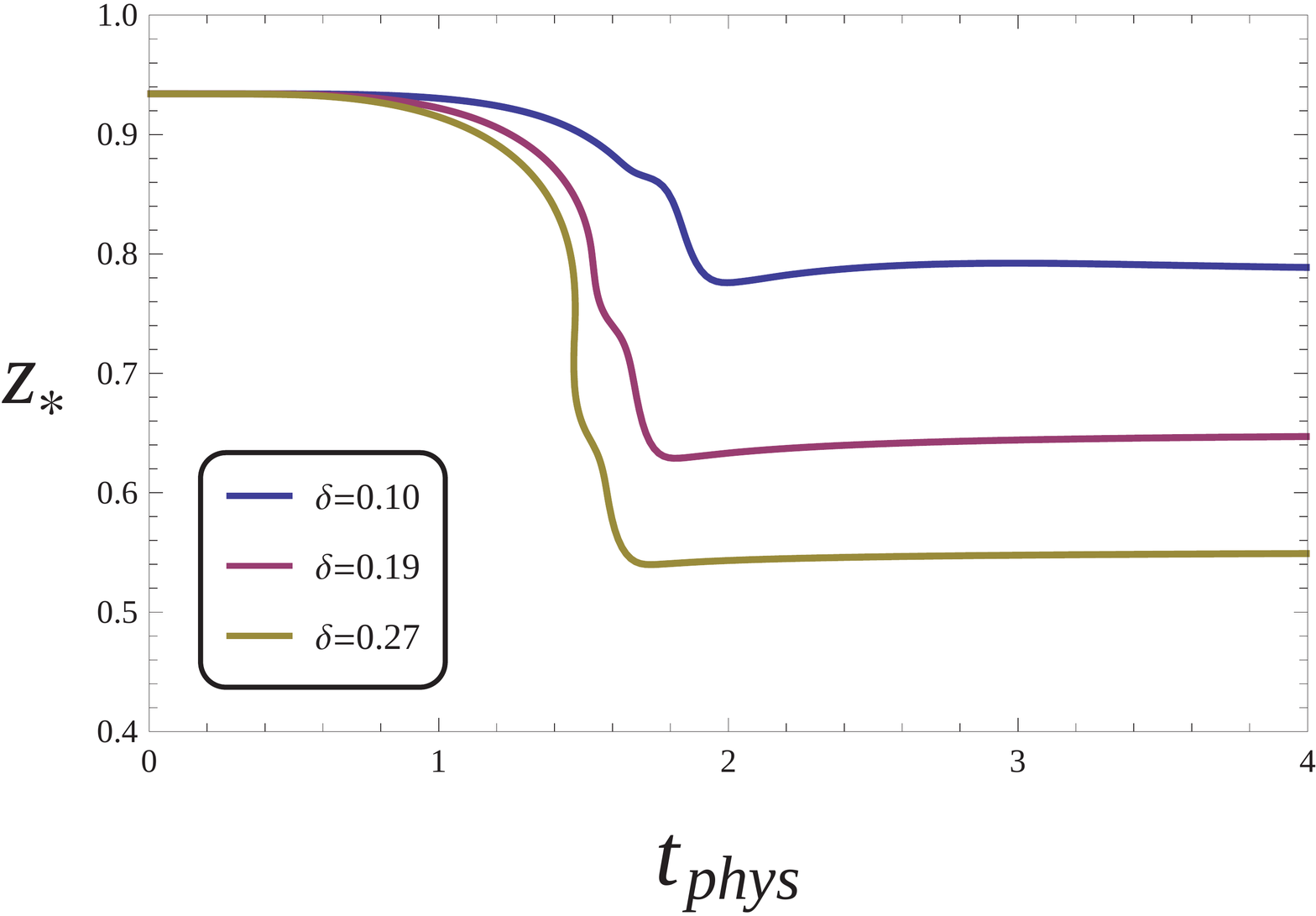}
\includegraphics[trim=0cm 0cm 0cm 0cm, clip=true, scale=0.26]{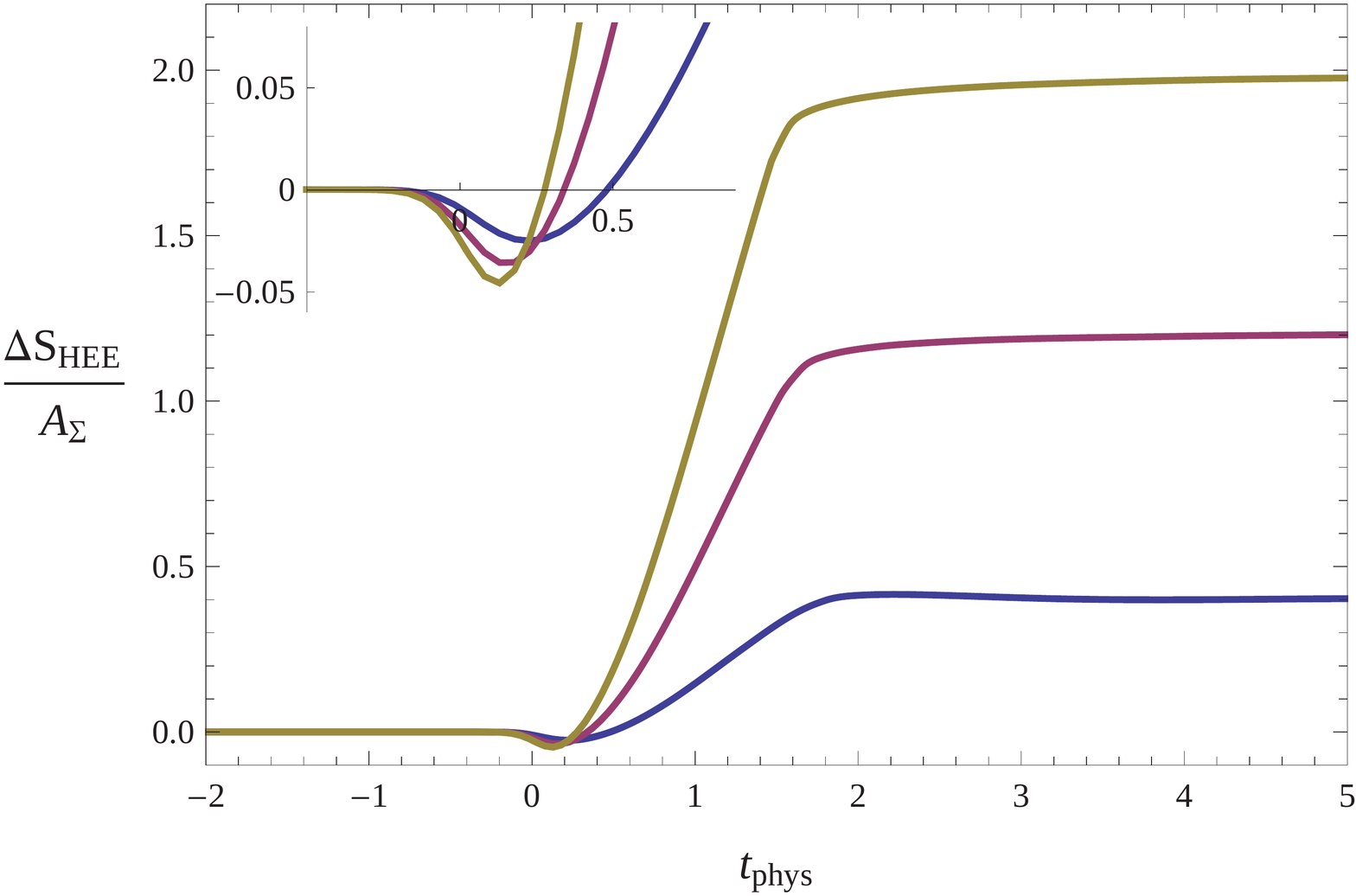}\\
$L=1.5$\\
\includegraphics[trim=0cm 0cm 0cm 0cm, clip=true, scale=0.26]{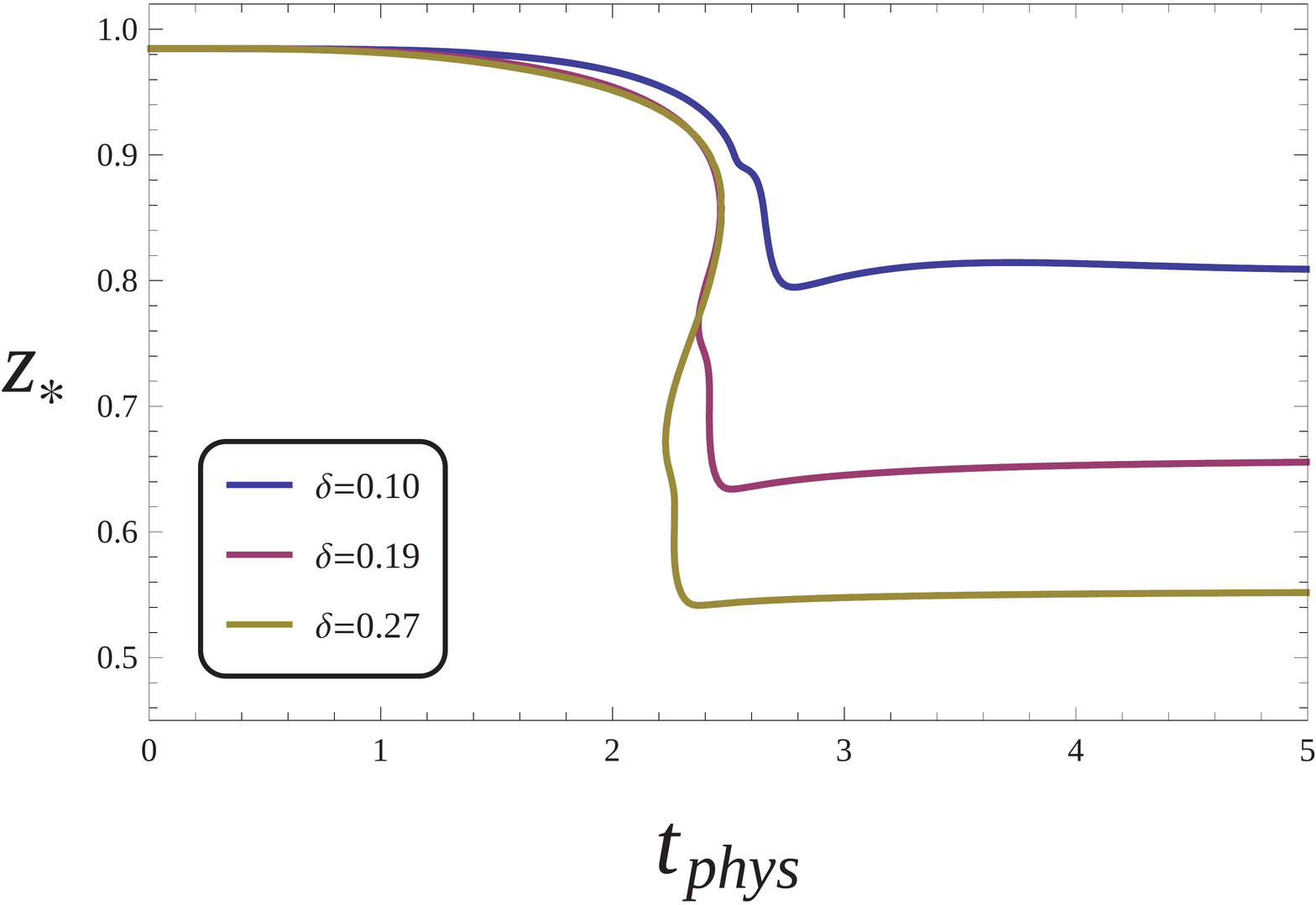}
\includegraphics[trim=0cm 0cm 0cm 0cm, clip=true, scale=0.26]{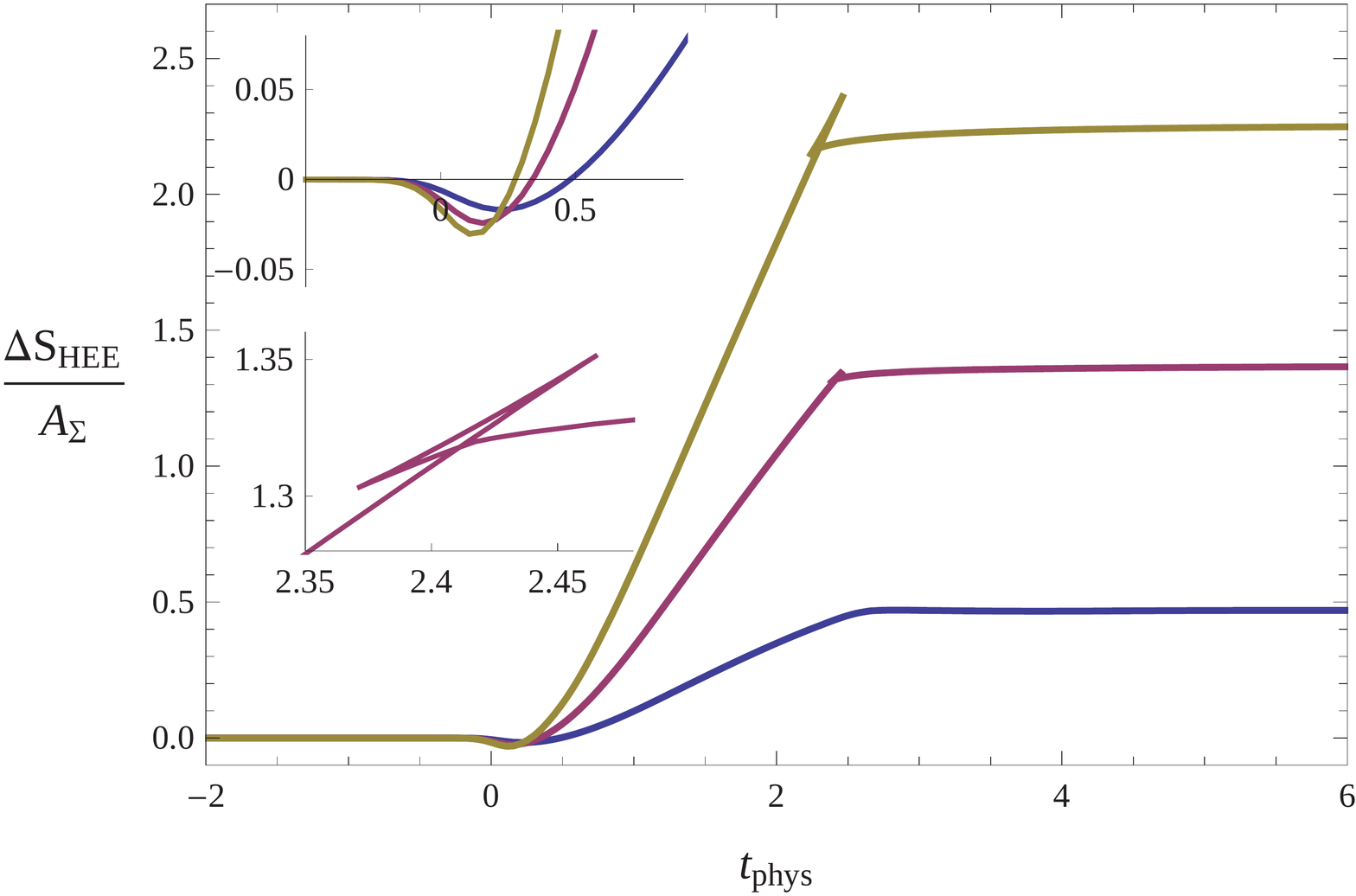}
\caption{\label{3Ls} (Left panels) Physical time evolution of the tip point; (Right panels) Physical time evolution of the density of the regularised entanglement entropy. The inset plots show a dip of $\Delta S_{\rm{HEE}}$ around $t_{\rm{phys}}=0$ (for $L=1.0$ and $L=1.5$), and a swallow tail behaviour at the saturation time (for $L=1.5$).}
\end{figure}

The Gaussian quench is centered at $t_{\rm{phys}}=0$, since the quench is from the source term $\psi_1$ on the boundary (see~\eqref{quench}). From the right panels of Fig.\ref{3Ls}, we see that near $t_{\rm{phys}}=0$, the entanglement entropy has a small dip due to this kind of Gaussian quench, in which $A_\Sigma=2LY$ is the area of the strip region.
For instance, from the inset plot of the right panel for $L=1.0$, we can find that the dip is deeper and steeper if the quench is stronger. This indicates that the sudden quench from boundary affects the HEE in an intricate way: the HEE will first decrease and then grow up after a short time.

This phenomenon is in contrast to the result in~\cite{Liu:2013iza} where there is a so called ``pre-local-equilibration" regime with quadratic growth in time instead of a dip. However, the authors of~\cite{Liu:2013iza} worked in the quench limit, taking the sourcing interval to zero. In particular, such a quench process is described by an infinitesimally thin shell of matter which collapses to form a black hole. In our present work, the quench is a Gaussian function with a finite sourcing interval (recall that we choose the time width of quench $\tau=0.5$). Compared with~\cite{Liu:2013iza}, it seems that the dip will disappear in the limit $\tau\rightarrow0$, for which we will leave for further study. Currently, the physics behind this dip form of HEE is still vague. It is really interesting to raise this problem in the hope that someone may come up with it in the future. The time evolution of HEE after a Gaussian quench was also studied in~\cite{Caceres:2014pda}. The authors considered a massless neutral scalar with a U(1) gauge field in the bulk and constructed a perturbative solution. At early time,
 the evolution of HEE also performed quadratic growth without any dip. This may suggest that the back reaction effect would be necessary to observe such dip.


The left panels of Fig.\ref{3Ls} show the tip points evolving with the physical time. We see that at $t_{\rm{phys}}=0$, all the points $z_*$ evolve smoothly, which means at this moment the quench still does not have significant impact on the tip points. { However, after some time $z_*$ shows a small ripple on it, which indicates that the effect of quench from the boundary has propagated to the tip points. For instance, for $L=1.0$ and $\delta=0.1$, the quench will affect the tip point at around $t_{\rm{phys}}=1.7$. Besides, we can also see that for a fixed length of the strip, {\it e.g.} $L=1.0$, stronger quench will affect $z_*$ more quickly, since the time the ripples come out are smaller for stronger quenches. Moreover, one can see that $z_*$ will go faster to the boundary (depart from $z=1$) if the quench becomes stronger.}

It also shows that at $t_{\rm{phys}}=0$, $z_*$ is smaller for the strips with shorter width, this is understandable from many previous literatures that the tip points will be much closer to the horizon if the width of the strip is longer. Even more, it will attach to the horizon if the width is long enough.

\subsection{Linear growth of HEE before equilibrium}

\begin{figure}[h]
\includegraphics[trim=0cm 0cm 0cm 0cm, clip=true, scale=0.26]{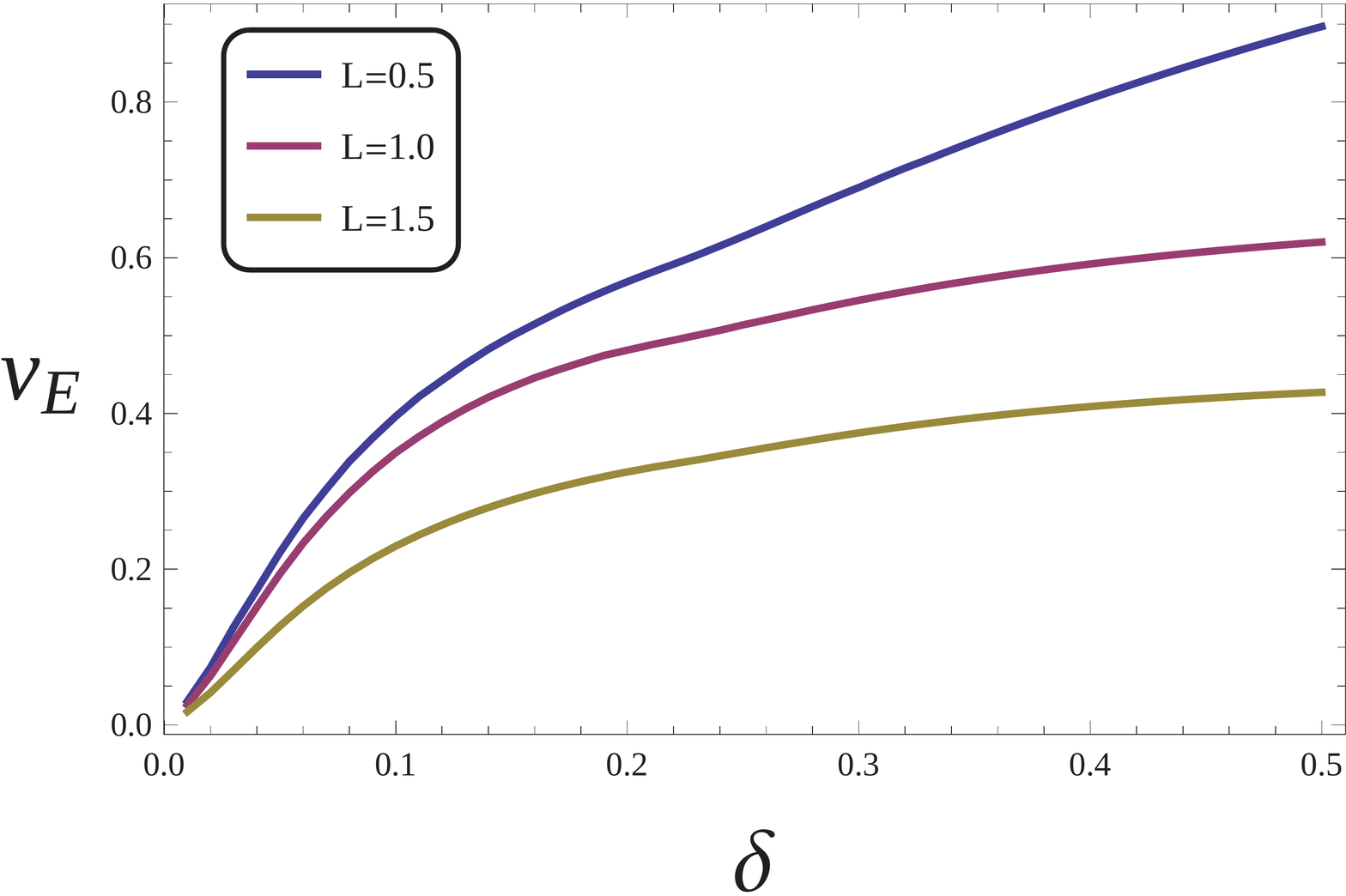}
\includegraphics[trim=0cm 0cm 0cm 0cm, clip=true, scale=0.26]{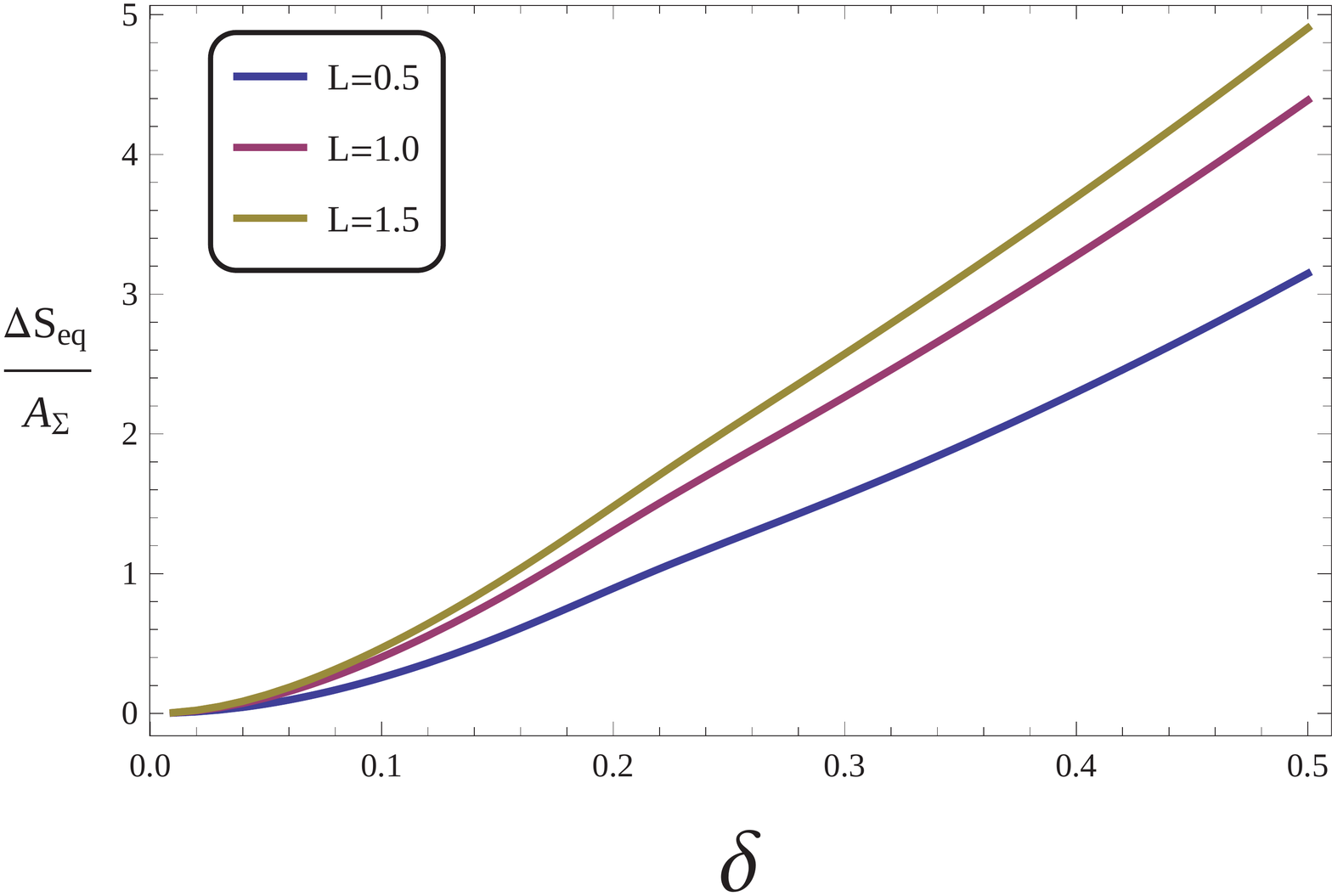}
\caption{\label{vEdelta} (Left) Linear growth velocity $v_E$ versus the quench strength $\delta$ for various $L$; (Right) Equilibrium entanglement entropy density versus quench strength $\delta$.}
\end{figure}

As time goes beyond the dip, HEE will perform a linear growth with $t_{\rm{phys}}$, which can be intuitively seen from the right panels of Fig.\ref{3Ls}. For a fixed width $L$, $\Delta S_{\rm{HEE}}$ will grow faster as the quench is stronger, which can be understood that stronger quench will change the system more abruptly. The authors of~\cite{Liu:2013iza,Liu:2013qca} defined a velocity $v_E$ in the linear growth regime as
\be
\frac{\Delta S_{\rm{HEE}}}{A_\Sigma}=s_{\rm{eq}}v_E t_{\rm{phys}},
\ee
in which $s_{\rm{eq}}$ is the equilibrium {\it thermal} entropy density, in our case $s_{\rm{eq}}=S(z_h)^2/z_h^2$ at the equilibrium state with $z_h$ the event horizon. It was postulated in~\cite{Liu:2013iza,Liu:2013qca} that $v_E$ should have an upper bound because of causality.\footnote{In the paper \cite{Liu:2013qca}, the authors argued that for $d=3$, $v_E\leq \sqrt{3}/2^{4/3}\approx 0.687$. In our paper we find that for larger width of the strip, say $L=1.0$ and $L=1.5$, $v_E$ satisfies this upper bound; However, for shorter width, say $L=0.5$, $v_E$ violates this bound. The reason may be that we used different setups for the geometric background and the quench. In  \cite{Liu:2013qca}, they adopted the geometric background as a pure AdS attached to a black hole along an in-falling collapsing null shell which is located at $t=0$. However, in our paper we adopted the superconducting black hole or hairy black hole as the background. } On the left panel of Fig.\ref{vEdelta}, we plot $v_E$ versus the quench strength for various strip lengths $L$. We can see that for every $L$, the velocity $v_E$ grows with the quench strength in certain parameter regimes. In particular, for $L=1.0$ and $L=1.5$ the velocity will finally tend to a finite value which is less than $1$ when $\delta$ is big enough. However, for $L=0.5$ it seems that the velocity does not tend to a fixed value in the regime of the parameters we choose in this paper. Maybe it will  tend to a fixed value when $\delta$ is large enough, however, due to the breakdown of the codes when $\delta$ is big enough, we cannot make any definite conclusion for $L=0.5$ at the present time.

\subsection{Saturation at equilibrium}

After the linear growth regime, $\Delta S_{\rm{HEE}}$ will generically saturate into an equilibrium state at a critical $t_{\rm{phys}}$ which can be seen from the right panels of Fig.\ref{3Ls}. This critical time can also be deduced form the left panel of Fig.\ref{3Ls} when the tip points $z_*$ tend to a flat value. For a fixed $L$, the equilibrium entanglement entropy is larger if the quench is stronger, which can be explained that stronger quench will pump more energy or degrees of freedom into the system. The continuous saturation can be found, for example in Fig.~\ref{3Ls} for $L=0.5$, in which at the critical time the derivative of $\Delta S_{\rm{HEE}}$ with respect to time is continuous.

An interesting thing is that the saturation of HEE may exhibit some swallow tails at the critical time for some parameter regimes, for example in the right bottom plot of Fig\ref{3Ls} for $\delta=0.19$ and $\delta=0.27$. It means that at the critical time there are multiple solutions of the surfaces that satisfy the EoMs \eqref{tz1} and \eqref{tz2} and the boundary conditions \eqref{bdy} and \eqref{bdy2}. However, for HEE we need to find a surface which has the minimal area. It is helpful to consult the left bottom plot of Fig.~\ref{3Ls} for the time evolution of the tip point $t_*$. We can see that for $\delta=0.27$, $z_*$ will go to the right as time evolves from $t_{\rm{phys}}=0$, then it will turn to left for a short while and then again turn right until saturating the equilibrium state. It can be found that from roughly $t_{\rm{phys}}=2.2$ to $t_{\rm{phys}}=2.5$ there exist multiple solutions to $z_*$. This is the reason why there is a swallow tail in the $\Delta S_{\rm{HEE}}$. This kind of swallow tail was also found in previous literatures, see for example~\cite{Albash:2010mv}.

On the right panel of Fig.\ref{vEdelta}, it shows the final equilibrium HEE $\Delta S_{\rm eq}$ versus the quench strength for strip subsystem with various widths. We can see that for a fixed $L$, the equilibrium HEE at late time will grow with the quench strength, which is consistent with the arguments we mentioned many times above; Moreover, for a fixed $\delta$, the equilibrium HEE is larger if the width $L$ is longer, which also matches the results we exhibited above. In addition, for a fixed $L$ the tip point $z_*$ at equilibrium will be closer to the boundary if the quench is stronger, which can be seen from the left panels of Fig.\ref{3Ls}.

\subsection{Apparent horizon, event horizon and the tip point $z_*$ }

\begin{figure}[h]
\includegraphics[trim=0cm 1.4cm 1.4cm 0cm, clip=true, scale=.35]{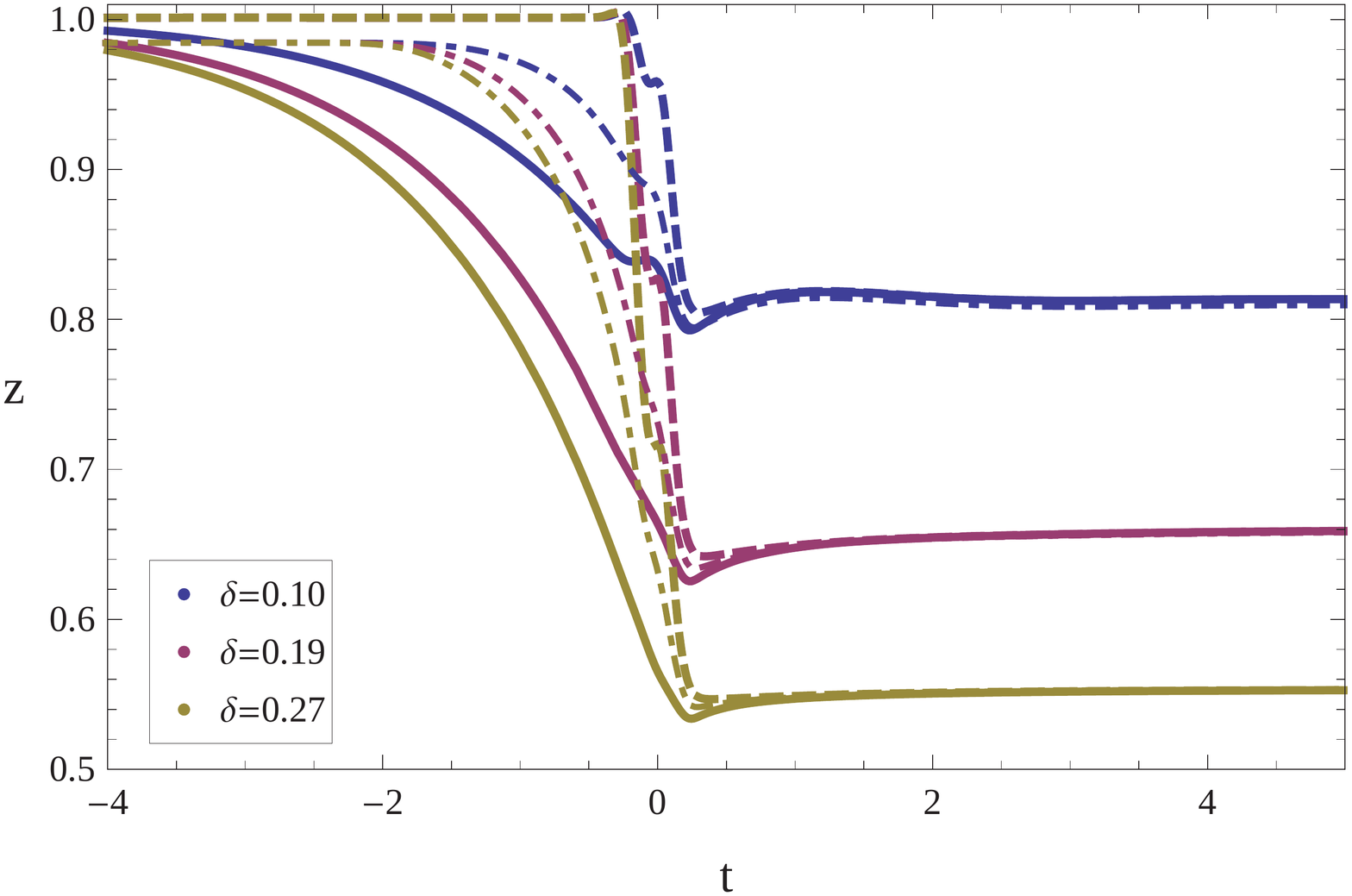}
\caption{\label{aeh} Time evolutions of apparent horizon, event horizon and the tip point $z_*$ for $L=1.5$ with respect to time $t$ for various quenches.  Dashed lines are the apparent horizons; Solid lines are the event horizons; Dash-dotted lines are the tip points $z_*$.}
\end{figure}

We plot the time evolution of apparent horizon, event horizon and the tip point $z_*$ in Fig.\ref{aeh} with respect to $t$. We can see that at the equilibrium states, the apparent horizon and the event horizon coincide as is expected. Tip point $z_*$ for $L=1.5$ will also meet the horizon at the late time. However, if $L$ is short, $z_*$ will not attach to the horizon at late time, which can be seen for example from the left top plot in Fig.\ref{3Ls} for $L=0.5$. There, for instance, when the quench strength $\delta=0.10$, the tip point $z_*$ at equilibrium will stay at around $z=0.64$, which is outside the horizon $z\approx0.81$.
At the equilibrium state in Fig.\ref{aeh}, we can see that when quench is stronger, the horizons will be much closer to the boundary. Meanwhile, the black hole surrounded by the horizon will be larger. This indicates that stronger quench will put more energy into the system and finally cross the horizon into black hole, thus makes black hole larger.

The information is intricate when $t<0$. It is interesting to see that for example $\delta=0.10$, the tip point $z_*$ may enter into the event horizon at a certain time and then finally meets the horizon at equilibrium time. However, we can also find that before equilibrium time, the location of apparent is always behind the event horizon, which gives a consistent check for our numerics.
\section{Phase transition from aspects of HEE}
\label{sect:pthee}
\begin{figure}[h]
\includegraphics[trim=0cm 0cm 0cm 0cm, clip=true, scale=.27]{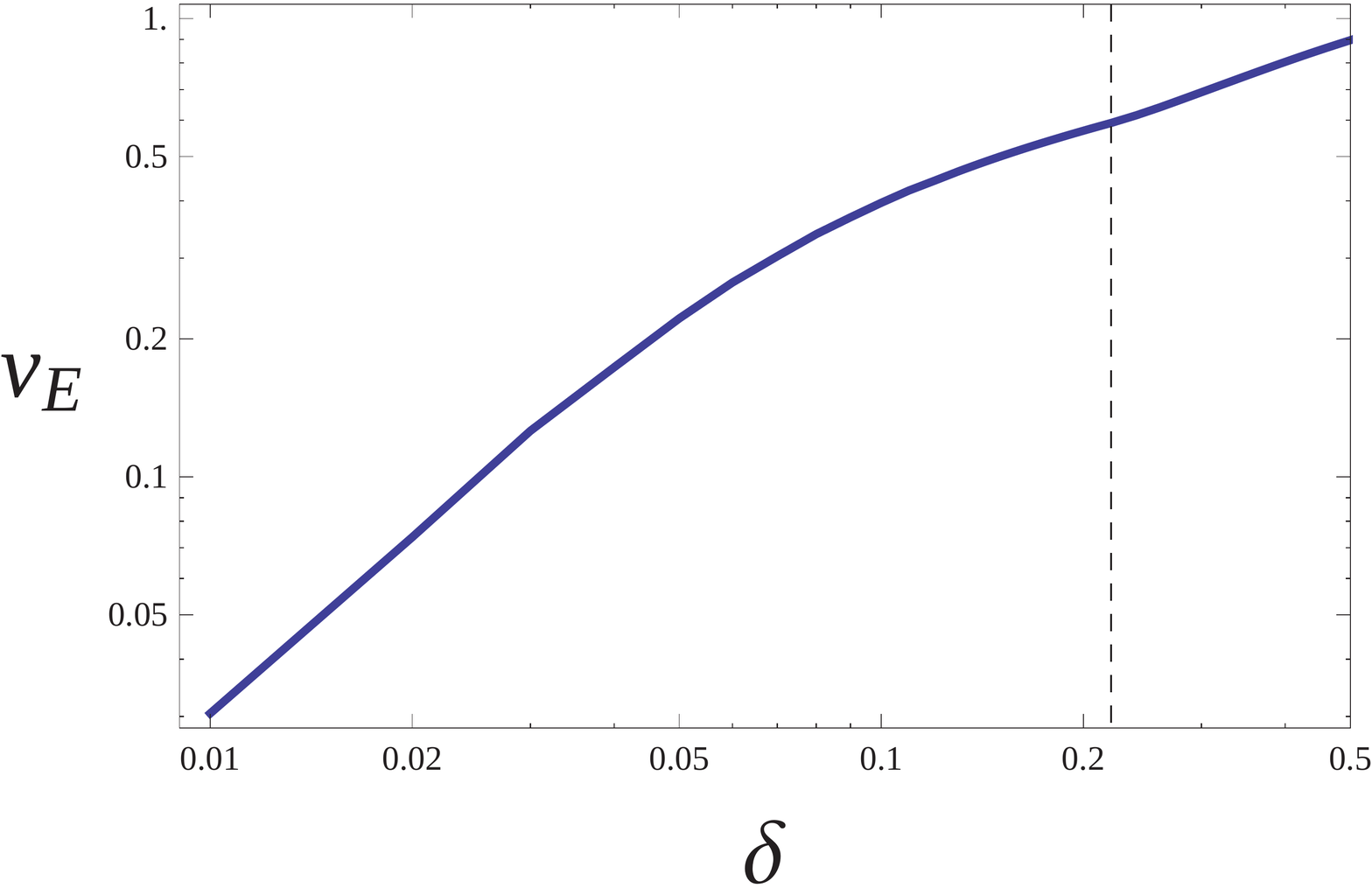}
\includegraphics[trim=0cm 0cm 0cm 1.5cm, clip=true, scale=.27]{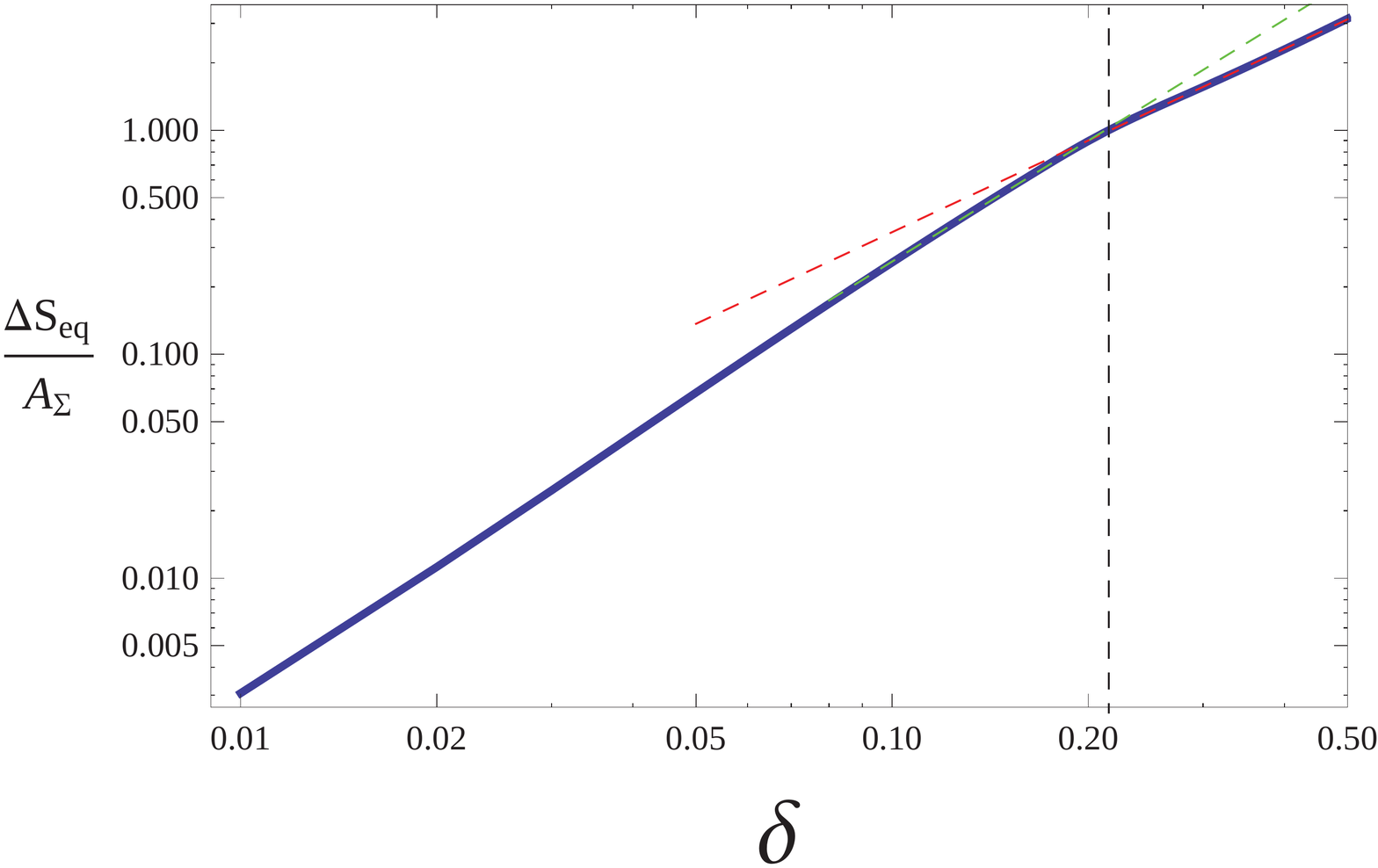}
\caption{\label{loglog} (Left) The log-log plot for the linear growth velocity $v_E$ versus quench strength $\delta$ for $L=0.5$; (Right) The log-log plot for regularized entanglement entropy density at equilibrium versus the quench strength $\delta$ for $L=0.5$. Red and green dashed lines are the fitted lines while the black dashed lines locate at $\delta\approx0.22$. }
\end{figure}

As we mentioned in Section \ref{sect:hsc} that there exists a phase transition around $\delta\approx0.22$, above which the hairy black hole at initial state will finally turn out to be a hairless black hole. We will try to uncover this phase transition by virtue of HEE. Actually, we do this from two points of view: One is from the behaviour of the linear growth velocity $v_E$, see the left panel of Fig.\ref{loglog}; the other one is from the behaviour of entanglement entropy density at equilibrium, see the right panel of Fig.\ref{loglog}. From Fig.\ref{vEdelta}, we can see that for $L=0.5$  each plot of $v_E$ or $\Delta S_{\rm eq}$ has a very tiny turning at $\delta\approx0.22$. In order to make this tiny effect more apparent, we plot $v_E$ and $\Delta S_{\rm eq}$ for $L=0.5$ in the log-log figures in Fig.\ref{loglog}.  From the left panel, we see that the scaling behaviour of $v_E$ is qualitatively different in the vicinity of $\delta\approx0.22$, which implies that $\delta\approx0.22$ is indeed a transition point. Similar behaviors can also be found for $L=1.0$ and $L=1.5$, which we do not show in this paper. A more precise or quantitative analysis of this phase transition point will be shown in the following in terms of $\Delta S_{\rm eq}$ at equilibrium.


The behaviour of $\Delta S_{\rm eq}$ can be found from the right panel of Fig.\ref{loglog}. We can see that the lines have distinct slopes before and after $\delta\approx0.22$, which also indicates that $\delta\approx0.22$ is a phase transition point. Green and red dashed lines are the fitted lines before and after the critical point respectively.  The linear behaviour in the log-log plot indicates that there is a power law scaling $\Delta S_{\rm eq}/A_{\Sigma}=\rho \delta^\zeta$ in the regular plot. The scaling behaviour $\Delta S_{\rm eq}/A_{\Sigma}=\rho \delta^\zeta$ for the green line is $\rho_1\sim16.098, \zeta_1=1.793$, while for the red line it is $\rho_2\sim8.023, \zeta_2=1.359$. We have also checked other strip width and found similar power law behaviour.

Therefore, we can  indeed deduce the phase transition point at $\delta\approx0.22$ from the behaviour of HEE. It is helpful to compare our HEE results with charged order parameter used to probe the dynamical phase transitions.
In the footnote~\ref{footnote} of Section~\ref{sect:hsc}, we mentioned that there also exists another ``dynamical" transition point at around $\delta_*\approx0.14$, at which the order parameter will behave from oscillatory damping to un-oscillatory damping but still has finite equilibrium values~\cite{Bhaseen:2012gg}.
However, from the analysis of HEE we cannot see such kind of transition at around $\delta_*\approx0.14$.
The time evolution of HEE under quench shows a common behaviour independent of the strength of quench:
it first develops a dip, then grows linearly and finally saturates. This seems reasonably as HEE is a non-local quantity. A priori, the HEE should behave more robust than local observables by changing external conditions, such as quench strength $\delta$. One possible reason is probably that the ``dynamical transition" near $\delta_*\approx0.14$ is not a phase transition but a smooth crossover, therefore HEE cannot probe this. Maybe there are still some hints in our data we ignored. It will be of great interest to investigate this issue in future.

\section{Conclusions and discussions}
\label{sect:con}

In this paper, we studied the dynamical evolution of entanglement entropy in a simple holographic superconductor model under an external Gaussian quench. The system is driven from the initial condensed state to a far-from-equilibrium regime, and finally equilibrates to a different equilibrium state which depends on the strength of quench~\cite{Bhaseen:2012gg}.  We calculated the HEE for a strip region during the whole dynamical process.

We found that the time evolution of HEE exhibits a common non-monotonic behaviour independent of the size of strip and quench strength.  There exists a small dip of HEE near the quench time. The depth and slope of the dip depended on the quench strength. In particular, if the quench is stronger the dip is deeper and steeper. Currently we do not know the exact physical meaning of the mysterious dip. Beyond the dip, HEE performs a famous linear growth with respect to the physical time for a boundary observer. By virtue of the formula in~\cite{Liu:2013iza}, we calculated the velocity $v_E$ of this linear growth and found that there is a bound for $v_E$ when the width of the strip is not much short. Besides, stronger quench would induce a faster linear growth of HEE, which indicated that stronger quench would change the system more abruptly. At a critical time, HEE would saturate into an equilibrium state. We found that if the width of the strip was large and the quench was strong, there exists a swallow tail at the critical time, otherwise the saturation at the critical time is continuous. In addition, when the quench is stronger, the final HEE at equilibrium is larger. More importantly, we found that the phase transition point at $\delta\approx0.22$ can be deduced by analyzing the linear growth velocity and equilibrium entropy density of HEE. However, The limitation by using HEE to probe the phase transition is that it cannot deduce the ``dynamical" transition point as we mentioned in the context. It will be of great interest to further study this in future time. Of course, extending this strip model to disc on the boundary is an obvious extension.

The initial condensed state  in our study is at a non-vanishing temperature. Depending on the strength of quench, this initial phase evolves to another equilibrium one with three distinct regimes for the behaviours of order parameters, which are precisely related to the spectrum of black hole quasi-normal modes \cite{Bhaseen:2012gg}. It is interesting to consider the initial state with vanishing temperature, which corresponds to the ground state of our superconducting system. If one turns on particular quench, the system will equilibrate at late time with a higher temperature. Form the gravity point of view, that is because the quench can inject energy into black hole. We expect that the dynamical evolution of the order parameter under quenches would not change qualitatively. However, the HEE may exhibit additional behaviour during the thermalization process from the ground state at vanishing temperature.

One can also generalise previous study to other gravity background, especially for AdS soliton which is confined and thus has been used to mimic insulator \cite{Nishioka:2009zj}. The behaviour of HEE in static case in such insulator/superconductor phase transition has been investigated in the literature. It was shown that the HEE exhibits distinct behaviour in AdS soliton compared to black hole case. It will be interesting to uncover the evolution of condensate and HEE after turning on quenches in this confined geometry.

\section*{Acknowledgement}
We would like to thank Rong-Gen Cai, Nava Gaddam, Hong Liu, Javier Martinez Magan, Keiju Murata, Julian Sonner, Phil Szepietowski and Toby Wiseman for helpful comments and discussions. We are also grateful to CERN for its hospitality and its partial support during the completion of this work. BHL was supported by the National Research Foundation of Korea (NRF) grant funded with grant number 2014R1A2A1A01002306;
LL was supported in part by European Union's Seventh Framework Programme under grant agreements (FP7-REGPOT-2012-2013-1) no 316165, the EU-Greece program ``Thales" MIS 375734 and was also co-financed by the European Union (European Social Fund, ESF) and Greek national funds through the Operational Program ``Education and Lifelong Learning" of the National Strategic Reference Framework (NSRF) under ``Funding of proposals that have received a positive evaluation in the 3rd and 4th Call of ERC Grant Schemes"; JRS was supported by the National Natural Science Foundation of China under Grant No. 11205058; HQZ was supported in part by the fund of Utrecht University budget associated to Gerard 't Hooft and the Young Scientists Fund of the National Natural Science Foundation of
China (No.11205097).

\end{CJK*}
\end{document}